\documentclass[a4paper,11pt]{article}

\usepackage{ifpdf}

\usepackage{epsfig,multicol,amsmath}
\usepackage[T1]{fontenc}

\usepackage{jheppub}
\usepackage{epstopdf}
\usepackage{amsmath}     
\usepackage{epsfig,epsf}
\usepackage{graphicx}
\usepackage{rotating}
\usepackage{verbatim}
\def\beq{\begin{equation}}
\def\eeq{\end{equation}}
\def\bea{\begin{eqnarray}}
\def\eea{\end{eqnarray}}

\def\eq#1{{Eq.~(\ref{#1})}}
\def\fig#1{{Fig.~\ref{#1}}}
\newcommand{\bas}{\bar{\alpha}_S}

\newcommand{\Lb}{\left(}
\newcommand{\Rb}{\right)}
\renewcommand{\vec}[1]{\boldsymbol{#1}}
\newcommand{\dif}{\mathrm{d}}

\parskip=\itemsep               

\newcommand{\nn}{\nonumber}
\newcommand{\D}{\partial}
\newcommand{\h}{\frac{1}{2}}

\title{CGC/saturation approach: a new impact-parameter dependent model.}
\author[a]{ Carlos Contreras,}
\author[a,b]{ Eugene ~ Levin}
\author[a]{  and Irina Potashnikova}

\affiliation[a]{Departemento de F\'isica, Universidad T\'ecnica Federico Santa Mar\'ia, and Centro Cient\'ifico-\\
Tecnol\'ogico de Valpara\'iso, Avda. Espana 1680, Casilla 110-V, Valpara\'iso, Chile}
\affiliation[b]{Department of Particle Physics, School of Physics and Astronomy,
Raymond and Beverly Sackler
 Faculty of Exact Science, Tel Aviv University, Tel Aviv, 69978, Israel}

\emailAdd{carlos.contreras@usm.cl}
\emailAdd{leving@post.tau.ac.il, eugeny.levin@usm.cl}
\emailAdd{irina.potashnikova@usm.cl}

\abstract{In this paper  we propose a new  impact-parameter dependent  CGC/saturation model. We introduce two new features in the model that make it
consistent with what we know theoretically about the deep inelastic scattering. They are:
the  use of  the exact form of the solution to the non-linear (BK) equation, whereas in all previous attempts only the form of  $r^2Q^2_s$ dependence, has been taken into account; and   the large impact parameter dependence, through the $b$-dependence of the saturation momentum which reproduce the correct behaviour of the amplitude at large impact parameters $b$ ($A \propto \exp\Lb - \mu b\Rb$) as well as at large momentum transferred $Q_T$ ($A $ decreases as a power of $Q_T$ as it follows from perturbative QCD). These improvement compared  to all previous attempts to build such models,  allows us to claim, that the experimental data are in accord with the prediction of CGC/saturation approach while previously, based on similar models, we could only conclude that the DIS data, perhaps,   can be described by introducing the shadowing corrections at small photon virtualities.
}

\keywords{  CGC/saturation approach, impact parameter dependence of the scattering amplitude, solution to non-linear equation, deep inelastic structure function, diffraction
 at high energies}
\begin{document}
\maketitle
\flushbottom

\pagestyle{empty}

\mbox{}

\pagestyle{plain}

\setcounter{page}{1}


\section{Introduction.}

In this paper we continue our  attempts to find an approach, based on Color Glass Condensate/saturation  effective theory  for high energy QCD (see  
Ref.\cite{KOLEB} for the review), which will satisfy the following requirements: it is simple but stems from the solution of the CGC/saturation equations. In other words, we would like to reduce the {\it ad hoc} or/and phenomenological ingredients to the minimum. We believe that the comparison of our model with a variety of  experimental data will allow us to demonstrate how the collective phenomena, that are incorporated in the CGC/saturation approach, are important  for understanding scattering processes at high energies.

For  comparison with  experimental data,  we choose the deep inelastic processes for which we have the CGC/saturation equations of Ref.\cite{JIMWLK} in a simple form of non-linear Balitsky-Kovchegov equation \cite{BK}. In addition to the simplicity of the theoretical interpretation of  the deep inelastic structure function $F_2$, these processes have been measured to  high accuracy \cite{HERA1,HERA2}, fitting to this observable allows us to determine  all phenomenological parameters that are  introduced in our model. However,  the theoretical interpretation of this observable should also include  the contribution of the colorless dipoles of large sizes,  which require a non-perturbative description in the framework of  QCD. In practice this  means that we are not able to apply our approach to low $Q^2$, where $Q$ is the photon virtuality. It is advantageous to compare $F_L$ with the data as  the main contribution to this structure function, stems from  dipoles of  size $1/Q$. However, only we have   scant data \cite{HERAFL1,HERAFL2} for   $F_L$.

After determining  the values of all phenomenological parameters of our approach from comparing to  the data on $F_2$, we calculate and  compare with the experimental data on open charm, and J/$\Psi$ meson diffractive production data \cite{HERASLOPE,HERAPSI,HERAZPHISL,HERAH1PSISL}. These data can be described within our theory with the same
accuracy as  $F_2$, as  the wave function of J/$\Psi$ meson is known. However,   other sources  of uncertainty such as the influence of the real part of the amplitude and/or the corrections that stem from   the skewedness effect,  due to the fact that the
gluons attached to the $q \bar{q}$ can carry different light-cone fractions $x$ and $x'$ of proton, have to be taken int account. They lead to uncertainties which are  not under full theoretical control. The same remarks  can be  made regarding  the virtual Compton reaction ($\gamma^* + p  \to \gamma + p$) \cite{HERAVIRTCOM} for which the wave function  is known to even better accuracy than for J/$\Psi$ production.

From the theoretical point of view,  the description of the diffractive production of $\phi$ and $\rho$ resonances  \cite{HERAH1PHIRHO} is less reliable,  since we do not have a  theoretical approach for the wave functions of these mesons, this in addition to all other uncertainties.

The main goal of  comparing with experimental data, is to show that the CGC/saturation approach is able to describe all these processes. This paper is neither the first  nor the last effort to build such a model based on the equations of CGC/saturation approach.  Numerous attempts have been undertaken over the past two decades (see Refs.\cite{SATMOD0,SATMOD1,BKL,SATMOD2,IIM,SATMOD3,SATMOD4,SATMOD5,SATMOD6,SATMOD7,SATMOD8,SATMOD9,SATMOD10, SATMOD11,SATMOD12,SATMOD13,SATMOD14,SATMOD15,SATMOD16,SATMOD17}) to build  such models. Therefore, we clarify, in the introduction,  the  aspects of our model  which are different.

First of all, it is necessary to construct a model, since  the CGC/saturation equations cannot reproduce the correct behavior of the scattering amplitude at large impact parameters (see Ref.\cite{KW,FIIM}).   Such failure leads, at least, to two conclusions: first, we cannot trust the solution of the CGC/saturation equations, and second, in  attempting to describe experimental data  we are doomed to build models to introduce the main features of the CGC/saturation approach.

We introduce the non-perturbative impact parameter behavior in the saturation moment, accordingly to the  spirit of  geometric scaling behavior of the scattering amplitude\cite{GS} and according to the semi-classical solution to the CGC/saturation equations  \cite{BKL}. Similar assumptions  for the non-perturbative $b$-behavior of the scattering amplitude, is typical most  models on the market (see Refs.\cite{SATMOD5,SATMOD6,SATMOD7,SATMOD8,SATMOD9,SATMOD12,SATMOD17}) and we refer only to the behavior at large $b$,  of the saturation scale $Q_s\Lb x, b\Rb$. We introduce $Q_s\Lb x, b\Rb \,\xrightarrow{b \gg 1/\mu}\,\exp\Lb - \mu b\Rb$ while  in all other  models $Q_s\Lb x, b\Rb \,\xrightarrow{b \gg 1/\mu}\,\exp\Lb - \mu^2 b^2\Rb$.
It should be stressed that the exponential decrease at large $b$, is in accord  the Froissart theorem \cite{FROI}. We would like to emphasize  that the $b$ dependence in our model reproduces the perturbative QCD prediction for the large values of the momentum transferred ($Q_T$)  leading to power-like behaviour of the scattering amplitude at large $Q_T$,  which cannot be reproduced for the Gaussian  $b$ distribution in other models.

The main difference stems from the way  we find the behavior of the dipole-nucleon scattering amplitude in the saturation region. In general, we follow the procedure suggested in Ref.\cite{IIM}:  we assume the geometric scaling behavior of the amplitude in the saturation region, and match the asymptotic behavior \cite{LETU} deep inside  the saturation region, with the behavior of this amplitude in the vicinity of the saturation scale \cite{IIML,MUT,KOLEB}. We also take into account, not only the correct behavior of the scattering amplitude inside the saturation region (see Ref.\cite{LETU}): viz. $N \propto \exp\Lb - A z^2\Rb $ with $z\,\,=\,\,\ln\Lb r^2 Q_s^2\Rb$ where $r$ is the size of the dipole and $Q_s$ is the saturation momentum, but the correct value of the coefficient $A$ \cite{LETU} and  we specify the procedure of matching at $z=0$. The procedure will be described below, and it is based on Ref.\cite{CLM}, to which we refer the reader for all details of this approach.

The natural question arises whether we introduce only  slight, cosmetic changes to  the models of Refs.\cite{SATMOD5,SATMOD6,SATMOD7,SATMOD8,SATMOD9,SATMOD12,SATMOD17} or the changes are  principal
in  nature. We  introduced in the model all theoretical information about the deep inelastic scattering, as, we believe, should be done in all phenomenological models, since the available experimental data are not sufficient to allow us to differentiate between theoretically correct and theoretically insufficient (or even wrong) approaches. It is enough to give an example of the deep inelastic scattering which can be described without any shadowing corrections (e.g.  Ref.\cite{LPDIS}). Therefore, we  need only  demonstrate that the theoretical changes that we introduce,
 are large enough to influence the description of the experimental data. This is the case since the correct $b$-behaviour leads to the $\ln^2(1/x)$ increase of the deep inelastic structure function $F_2$ ,while the correct behaviour at large $z$ determines the coefficient in front of $\ln(1/x)$ in $F_2$. Our numerical estimates show that both factors are essential in the description of the HERA data, and  become even more significant at higher energies.

 As will be seen from the body of the paper, our model provides a good description of the experimental data. The fact that this description does not appear  better than in the other insufficient  models on the market, allows us to claim that the CGC/saturation approach does not contradict the available experimental data. On the other hand we learned only that
 shadowing corrections are, perhaps, needed to describe the experimental data   from comparison with the experiment all models of Refs.\cite{SATMOD0,SATMOD1,BKL,SATMOD2,IIM,SATMOD3,SATMOD4,SATMOD5,SATMOD6,SATMOD7,SATMOD8,SATMOD9,SATMOD10, SATMOD11,SATMOD12,SATMOD13,SATMOD14,SATMOD15,SATMOD16,SATMOD17}.
 

\section{Theoretical input}
\subsection{General formula}
\begin{figure}[ht]
    \centering
  \leavevmode
      \includegraphics[width=10cm]{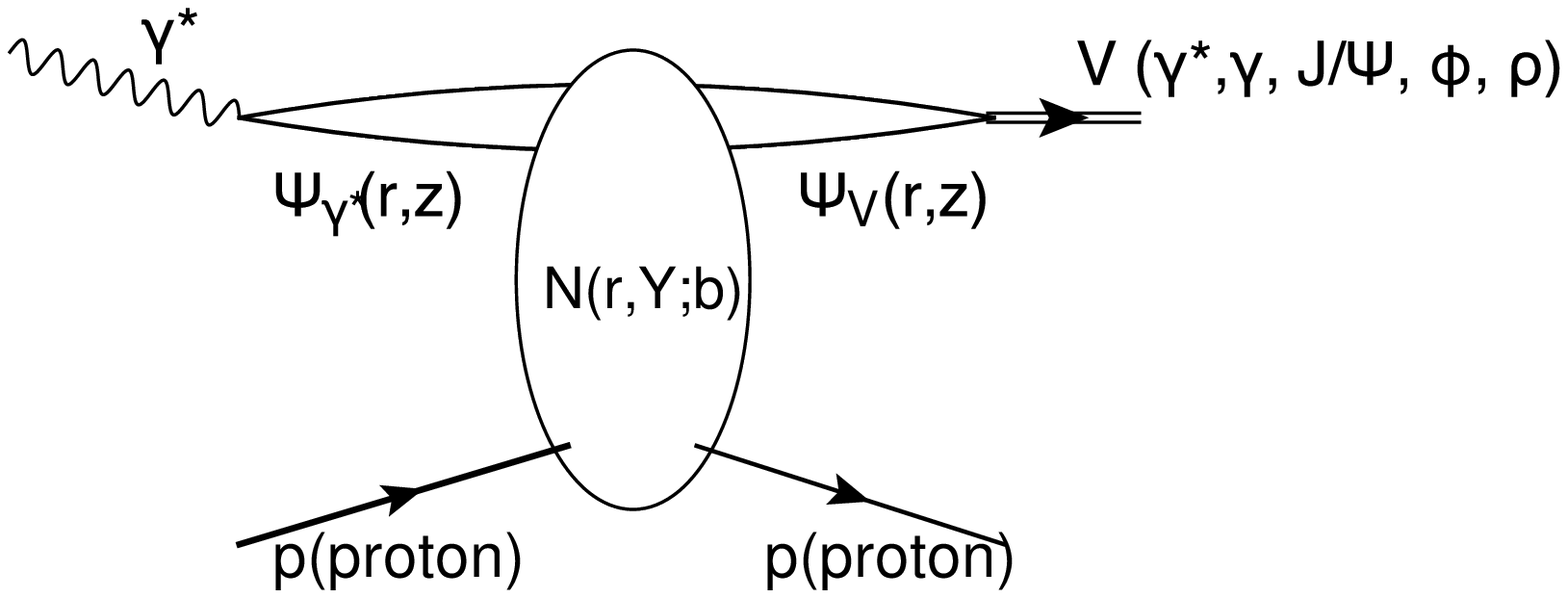}  
      \caption{The graphic representation of \protect\eq{FORMULA} for the scattering amplitude. $Y = \ln\Lb 1/x_{Bj}\Rb$ and $r$ is the size of the interacting dipole.   $z$ is the fraction of energy that is carried by one quark. $b$ is the impact parameter of the scattering amplitude}
\label{gen}
   \end{figure}

The general formula for  deep inelastic processes takes the form (see \fig{gen} and Ref. \cite{KOLEB} for the review and references therein)

\beq\label{FORMULA}
N\Lb Q, Y; b\Rb \,\,=\,\,\int \frac{d^2 r}{4\,\pi} \int^1_0 d z \,\Psi_{\gamma^*}\Lb Q, r, z\Rb \,N\Lb r, Y; b\Rb\,\Psi^*_V\Lb r,z\Rb
\eeq
where $Y \,=\,\ln\Lb 1/x_{Bj}\Rb$ and $x_{Bj}$ is the Bjorken $x$. $z$ is the fraction of energy carried by quark.
$Q$ is the photon virtuality. $b$ is the impact parameter of the scattering amplitude.

\eq{FORMULA} splits the calculation of the scattering amplitude into two stages: calculation of the wave functions, and estimates of the dipole scattering amplitude.

\subsection{Dipole-nucleon scattering amplitude}
We follow the general plan of Ref.\cite{IIM},  searching for a  solution to the CGC/saturation equation,  by matching the solution in the vicinity of the saturation scale, with the solution deep inside the saturation domain. In both these regions we have  analytical solutions. 


\subsubsection{Vicinity of the saturation scale}
Dipoles of sizes $r$ are such, that $r^2\,Q^2_s\Lb Y,b\Rb \,\to\,1$ the scattering \cite{IIML,MUT,KOLEB}
\beq \label{N1}
N\Lb r, Y; b\Rb\,\,=\,\,N_0 \Lb r^2 \,Q^2_s\Lb b \Rb \Rb^{1 - \gamma_{cr}}
\eeq
where  $\gamma_{cr}$ is the solution of the equation
\beq \label{N2}
\frac{\chi\Lb \gamma_{cr}\Rb}{1 -   \gamma_{cr}}\,\,=\,\Big{|} \frac{d \chi\Lb \gamma_{cr}\Rb}{d \gamma_{cr}}\Big{|}
\eeq
In \eq{N2}  $\chi\Lb \gamma\Rb$ is the BFKL kernel that takes the form
\beq \label{N3}
\chi\Lb \gamma \Rb\,=\,2\,\psi\Lb 1 \Rb \,-\, \psi\Lb \gamma\Rb \,-\, \psi\Lb 1 -  \gamma\Rb
\eeq
where $\psi\Lb z \Rb$ is the digamma function. 

The scattering amplitude of \eq{N2} shows the geometric scaling behaviour  \cite{GS}. In Ref. \cite{IIML} it  is shown that the first corrections due to a violation of the geometric scaling behavior  can be taken into account by replacing $1 - \gamma_{cr}$ in \eq{N1}
by the following expression
\beq \label{N4}
1 \,-\,\gamma_{cr}\,\,\to\,\,1 \,-\,\gamma_{cr}\,\,-\,\,\frac{1}{2\,\kappa\,\lambda\,Y} \,\ln\ \Lb r^2 \,Q^2_s\Lb b \Rb \Rb
\eeq
where $\lambda = \bas \Lb \chi\Lb \gamma_{cr}\Rb/\Lb 1 - \gamma_{cr}\Rb\Rb$ and $\kappa \,=\,\chi''\Lb \gamma_{cr}\Rb/\chi'\Lb \gamma_{cr}\Rb$.


\subsubsection{Saturation region}

The new ingredient in  our model is the way we  found the solution to the Balitsky-Kovchegov (BK) equation \cite{BK} in the saturation  region where $r^2\,Q^2_s\Lb Y,b\Rb \,>\,1$.  In this region the scattering amplitude demonstrates  geometric scaling behavior,  being a function of one  variable 
\beq \label{z}
z\,\,=\,\,\ln\Lb r^2\,Q^2_s\Lb Y,b\Rb \Rb\,\,=\,\,\ln\Lb r^2 Q^2\Lb Y=Y_0, b\Rb\Rb\,\,+\,\,\lambda \Lb Y - Y_0\Rb\,\,=\,\,\xi\,\,+\,\,\lambda\,\Lb Y - Y_0\Rb
\eeq
since  the saturation scale behaves as
\beq \label{QS}
Q^2_s\Lb Y, b \Rb\,\,=\,\,Q^2\Lb Y = Y_0,b\Rb e^{\lambda\,\Lb Y - Y_0\Rb}\,\,=\,\,Q^2\Lb Y = Y_0,b\Rb \Lb \frac{x_0}{x}\Rb^\lambda
\eeq
where $Y = \ln(1/x)$,  $x$ is the Bjorken energy variable which is equal to $x = Q^2/s$  for deep inelastic scattering. $Q$ is the photon virtuality and $s$ is the energy squared for the process.

Deep in the saturation region ($z \,\,\gg\,\,1, r^2\,Q^2_s\Lb Y, b\Rb\,\gg\,1$)  the solution to the non-linear BK equation is known (see  Ref.\cite{LETU}). In Refs.\cite{CLM} the next order correction to asymptotic behavior at large $z$ has been calculated. In this paper we use the following expression taken from Ref.\cite{CLM} 
\bea \label{MF}
N^{z \,\gg\,1} \Lb z \Rb &\,=\,&1 - 2\,A e^{- {\cal Z}}\,\,-\,\sqrt{2 \lambda} \,A^2\,\frac{1}{\sqrt{{\cal Z} }}e^{- 2 {\cal Z}}\,\,+\,\,{\cal O}\Lb e^{-3 {\cal Z}}\Rb\\
{\cal Z}&=&\frac{\Lb z -  \h A \sqrt{\lambda \pi/2} - 2\psi(1)\Rb^2}{2 \lambda}\nn
\eea
where $\psi(x)$ is the digamma function (see Ref.\cite{RY} formula {\bf 8.360 - 8.367}).

The second term in \eq{MF} is the solution given in  Ref.\cite{LETU} in which the theoretically unknown constant $A$ is introduced both as the coefficient in front and as correction to the argument. The third term is the next order correction  at large $z$.

For $z > 0 $ but $z\, \ll\,1$,  the solution to BK equation has been found in Ref.\cite{CLM} and it takes the form

 \beq \label{MF1} 
 N^{0 <z \ll 1}\Lb z \Rb \,\,=\,\,-2\, \frac{ \lambda \Lb 1 - \gamma_{cr}\Rb^2 \Lb N_0 - \lambda \Lb 1 - \gamma_{cr}\Rb \Rb \,N_0\, e^{ - \Lb 1 - \gamma_{cr}\Rb \,z}}{\Lb N_0 \,\,-\,\,\Lb N_0\,-\, \lambda  \Lb 1 - \gamma_{cr}\Rb\Rb e^{ - \Lb 1 - \gamma_{cr}\Rb \,z}\Rb^2}
 \eeq
where $N_0$ is the value of the scattering amplitude at $z = 0$.

We match these two solution at $z = z_{m}$ where
\beq \label{MC}
 N^{0 < z \ll 1}\Lb z = z_m \Rb \,=\, N^{z \gg  1}\Lb z = z_m \Rb;~~~~~~~~~~~ \frac{d N^{0 <z \ll 1}\Lb z = z_m \Rb}{d z_m} \,=\, \frac{d N^{z \gg  1}\Lb z = z_m \Rb}{ d z_m};~ 
 \eeq
 It has been demonstrated in Ref.\cite{CLM} that we  find $z_m \approx1$.
  \subsubsection{Estimates for the parameters of the scattering amplitude} 
 The entries  in \eq{N1}, \eq{MF} and \eq{MF1} can be divided in two groups: parameters that can be evaluated in the leading log(1/x) approximation of perturbative QCD, and  phenomenological parameters that can be found only  by  comparing with the experimental data. The value of $\gamma_{cr}$ is determined by \eq{N2} and it is equal to 0.37. The energy dependence of the saturation scale is given by 
 \beq \label{LAM}
 \lambda = \bas\chi\Lb \gamma_{cr}\Rb/(1 - \gamma_{cr}). 
 \eeq
 In \eq{N4}  $\kappa \,=\,\chi''\Lb \gamma_{cr}\Rb/\chi'\Lb \gamma_{cr}\Rb$ in the leading order of perturbative QCD(LO). In our procedure of fitting we consider the value of $\lambda$ as a parameter of the fit since we do not know the value of $\bas$  and the next-to-leading order corrections (NLO) in perturbative QCD change considerably the value of $\lambda$ \cite{MUT,KMRS} making it in 3-4 times less than the LO estimates. The value of $\lambda$ is determined by the linear BFKL equation which is known in the NLO, and therefore, we consider the
 estimates of $\lambda$ as reliable, which have to be taken into account in spite of uncertainties in the value. $\lambda \approx 0.2$ which we obtain from the fit is close to the NLO estimates.
 
 In principle  the value of $N_0$ can be calculated   using the linear evolution equation with  the initial conditions. However, it  depends on the phenomenological parameters of this initial condition. We choose  to  use  $N_0$ as the parameter of the fit.
 
 The impact parameter dependence of the saturation scale as we have discussed, is the pure phenomenological input, which can only  be estimated in non-perturbative QCD. For  $Q_s\Lb Y = Y_0,b\Rb$ we use the following expression
 \beq \label{QSB}
   Q^2_s\Lb Y = Y_0,b\Rb\,\,=\,\,m^2\,S\Lb b \Rb\,\,=\,\, m^2 \Lb m\,b\,K_1\Lb m\, b\Rb\Rb^{1/(1 - \gamma_{cr})}
   \eeq
 $m$ must  be determined from the fit. We differ  from other models in that \eq{QSB} leads to\\ $   Q^2_s\Lb Y = Y_0,b\Rb\,\xrightarrow{m b \,\gg\,1}
 \exp\Lb - m b/(1 - \gamma_{cr})\Rb$  providing the correct large $b$ behavior of the scattering amplitude.  In other models (see Refs.\cite{SATMOD5,SATMOD6,SATMOD7,SATMOD8,SATMOD9,SATMOD12,SATMOD17})   $   Q^2_s\Lb Y = Y_0,b\Rb\,\propto \exp\Lb- b^2/B\Rb$.
 
\subsection{Wave functions}
The wave function in the master equation (see \eq{FORMULA}) is the main source of theoretical uncertainties: even in the case of  deep inelastic processes,  we can trust the wave function of perturbative QCD only, at rather large values of $Q^2 \,\geq\,Q^2_0$ with $Q^2_0 \approx 0.7 GeV^2$ (see Ref. \cite{GLMTC}). The expression for $(\Psi^*\Psi)^{\gamma^*} \equiv \Psi_{\gamma^*}\Lb Q, r, z\Rb \,\Psi_{\gamma^*}\Lb  Q, r,z\Rb$ is well known (see Ref.\cite{KOLEB} and references therein)
\begin{align}
  (\Psi^*\Psi)_{T}^{\gamma^*} &=
   \frac{2N_c}{\pi}\alpha_{\mathrm{em}}\sum_f e_f^2\left\{\left[z^2+(1-z)^2\right]\epsilon^2 K_1^2(\epsilon r) + m_f^2 K_0^2(\epsilon r)\right\},\label{WFDIST}   
  \\
  (\Psi^*\Psi)_{L}^{\gamma^*}&
  = \frac{8N_c}{\pi}\alpha_{\mathrm{em}} \sum_f e_f^2 Q^2 z^2(1-z)^2 K_0^2(\epsilon r),
\label{WFDISL}
\end{align}
where T(L) denotes the polarization of the photon and $f$ is the flavours of the quarks. $\epsilon^2\,\,=\,\,m^2_f\,\,+\,\,Q^2 z (1 - z)$.

 In addition to the total DIS cross section, the wave function is known theoretically  for deeply virtual Compton scattering (DVCS). For this process 
 $(\Psi^*\Psi)^{DVCS} \equiv \Psi_{\gamma}\Lb r, z\Rb \,\Psi_{\gamma^*}\Lb  Q, r,z\Rb$ is equal to
 \begin{equation}
  (\Psi_\gamma^*\Psi)_{T}^{DVCS} = \frac{2N_c}{\pi}\alpha_{\mathrm{em}}\sum_f e_f^2\left\{\left[z^2+(1-z)^2\right]\epsilon K_1(\epsilon r) m_f K_1(m_f r)+ m_f^2 K_0(\epsilon r) K_0(m_f r)\right\}.
  \label{DVCS}
\end{equation} 
 Since in DVCS the real photon is produced, so   its polarization can be only transverse,  as  seen from \eq{DVCS}.
 
For  vector meson diffractive production,  the wave functions are known only for the mesons that are constituted  of heavy quarks. The most popular example is the J/$\Psi$ diffractive production, but we have to bear in mind, that the mass of the charm quark is not very large,  and corrections can be essential. For all other mesons the confinement corrections are large and the form of the wave functions motivated by the heavy quark mesons, can be considered only as a pure phenomenological conjecture. We use the following form of the overlaps integrals  $(\Psi^{*}_V\Psi)_{T,L} \equiv \Psi_{\gamma^*}\Lb Q, r, z\Rb \,\Psi^{V}_{T,L}\Lb  r,z\Rb$ taking them from Ref.\cite{SATMOD6}
\begin{align}
  (\Psi_V^*\Psi)_{T} &= \hat{e}_f e\, \frac{N_c}{\pi z(1-z)} \,
  \left\{m_f^2 K_0(\epsilon r)\phi_T(r,z) - \left[z^2+(1-z)^2\right]\epsilon K_1(\epsilon r) \partial_r \phi_T(r,z)\right\},
  \label{VT}
  \\
  (\Psi_V^*\Psi)_{L} &=  \, \hat{e}_f e \, \frac{N_c}{\pi}\,
  2Qz(1-z)\,K_0(\epsilon r)\,
  \left[M_V\phi_L(r,z)+ \frac{m_f^2 - \nabla_r^2}{M_Vz(1-z)}
    \phi_L(r,z)\right],
  \label{VL}\\
  \phi_{T,L}(r,z) &= \mathcal{N}_{T,L} z(1-z)
  \exp\left(-\frac{m_f^2 \mathcal{R}^2}{8z(1-z)} - \frac{2z(1-z)r^2}{\mathcal{R}^2} + \frac{m_f^2\mathcal{R}^2}{2}\right)
  \label{philt}
\end{align}
where the effective charge $\hat{e}_f=2/3$, $1/3$, or $1/\sqrt{2}$, for $J/\psi$, $\phi$, or $\rho$ mesons respectively. 
$m_f = 140 \,MeV$ is  used in all above equations and the values of $ \mathcal{N}_{T,L}$ and $\mathcal{R}$
are taken from Table 2 in Ref. \cite{SATMOD3}.

\subsection{Other theoretical uncertainties}
\subsubsection{Real part of the scattering amplitude}
The master equation (see \eq{FORMULA}) is derived assuming that the scattering amplitude is purely imaginary. To account  for the real part of the amplitude,   we need to multiply the differential cross section calculated by using \eq{FORMULA} by factor  $ 1 + \rho^2$ where $\rho$ is the ratio of real to imaginary parts of the scattering amplitude $A$. In principle, we can calculate the real part of the amplitude if we know the imaginary part using the dispersion relation. For large values of energy this calculation  simplifies to
\beq \label{RHO}
\rho \,\,=\,\,\frac{\pi}{2} \frac{d \ln\Lb N\Rb}{d \ln \Lb 1/x\Rb}
\eeq
for a  more general formula for the amplitudes that show Regge-type behavior at high energies (see Ref.\cite{GLMIP} and references therein)
\begin{equation}\label{RHOF}
  \rho = \tan(\pi\beta/2), \quad\text{with}\quad \beta  \equiv \frac{\partial\ln\left(N_{T,L}^{\gamma^* p\rightarrow V p}\right)}{\partial\ln(1/x)}.
\end{equation}

\subsubsection{Off diagonal contributions}
For vector meson production or DVCS, we  need to modify the main formula of \eq{FORMULA}, to 
account for  the fact  that the virtual photon, and the produced vector mesons carry 
different fractions $x$ and $x^\prime$ of the proton's (light-cone) momentum.  In the leading $\ln(1/x)$ limit, this imbalance (skewed effect) does not affect the scattering amplitude, but it has to be taken into account 
in the limit that $x^\prime \ll x \ll 1$.  Two include the skewed effect we multiply the amplitude by the factor $R_G$
given by \cite{SGMR}

\begin{equation} \label{RG}
  R_g(\beta) = \frac{2^{2\beta+3}}{\sqrt{\pi}}\frac{\Gamma(\beta+5/2)}{\Gamma(\beta+4)}
\end{equation}
where $\beta$ is given by \eq{RHOF}.
\section{Fitting $F_2$ and values of the parameters}
The most accurate  experimental data  available  are on the deep inelastic structure function $F_2$ \cite{HERA1}. Our strategy is to  determine all phenomenological parameters by  fitting to these data. Other data we will use to illustrate that our model can be used  as a  simple and convenient   tool in describing a variety of processes. As has been mentioned, we can trust our model in the restricted kinematic region, which we choose in the following way: $ 0.85\,GeV^2 \leq Q^2\leq 60\,GeV^2$ and  $x \,\leq \,0.01$. The lower limit of $Q^2$ stems from non-perturbative correction to the wave  function of the virtual photon, while the upper limit is originated from the restriction $x\,\leq\,0.01$.


\begin{table}[ht]
\begin{tabular}{||l|l|l|l|l|l|l|l|l||}
\hline
\hline
$\lambda $ & $N_0$ & m ($GeV$)& $Q^2_0$ ($GeV^2$) & $m_u $(MeV) &  $m_d $(MeV) &  $m_s $(MeV)&  $m_c $(GeV)  &$\chi^2/d.o.f.$ \\
\hline
0.197& 0.34 & 0.75 & 0.145& 2.3 &4.8& 95&1.4& 178/155 =1.15\\
\hline
0.184 & 0.46 & 0.75 & 0.118& 140 &140 &140& 1.4 & 176/154 = 1.14\\
\hline\hline
\end{tabular}
\caption{Fitted parameters of the model.   $Q^2_0 \,=\,m^2\, x^\lambda_0$.}
\label{t1}
\end{table}
 
We have the following fitting parameters: $\lambda$ for $x$-dependence of the saturation momentum, $N_0$  for the value of the scattering amplitude at $r^2 \,Q^2_s =1$, $m$ for the impact dependence of the saturation scale and $x_0$ or $Q^2_0 = m^2 x^\lambda_0$ for the value of the saturation scale. Masses of the quark we do not regard  as fitting parameters and consider two sets of these masses. In the first set we take the current masses (see the first row of Table 1) and we consider this as the most reliable fit based on the consistent theoretical approach.We  also make  a fit  putting all masses of light quarks (second row of Table 1) to be equal to 140 MeV. We view this mass as a typical infra-red cutoff , which was used in the wave functions of the produced mesons ($m_f $= 140 MeV in \eq{VT} and \eq{VL}).

\begin{figure}[h]
    \centering
  \leavevmode
      \includegraphics[width=14cm]{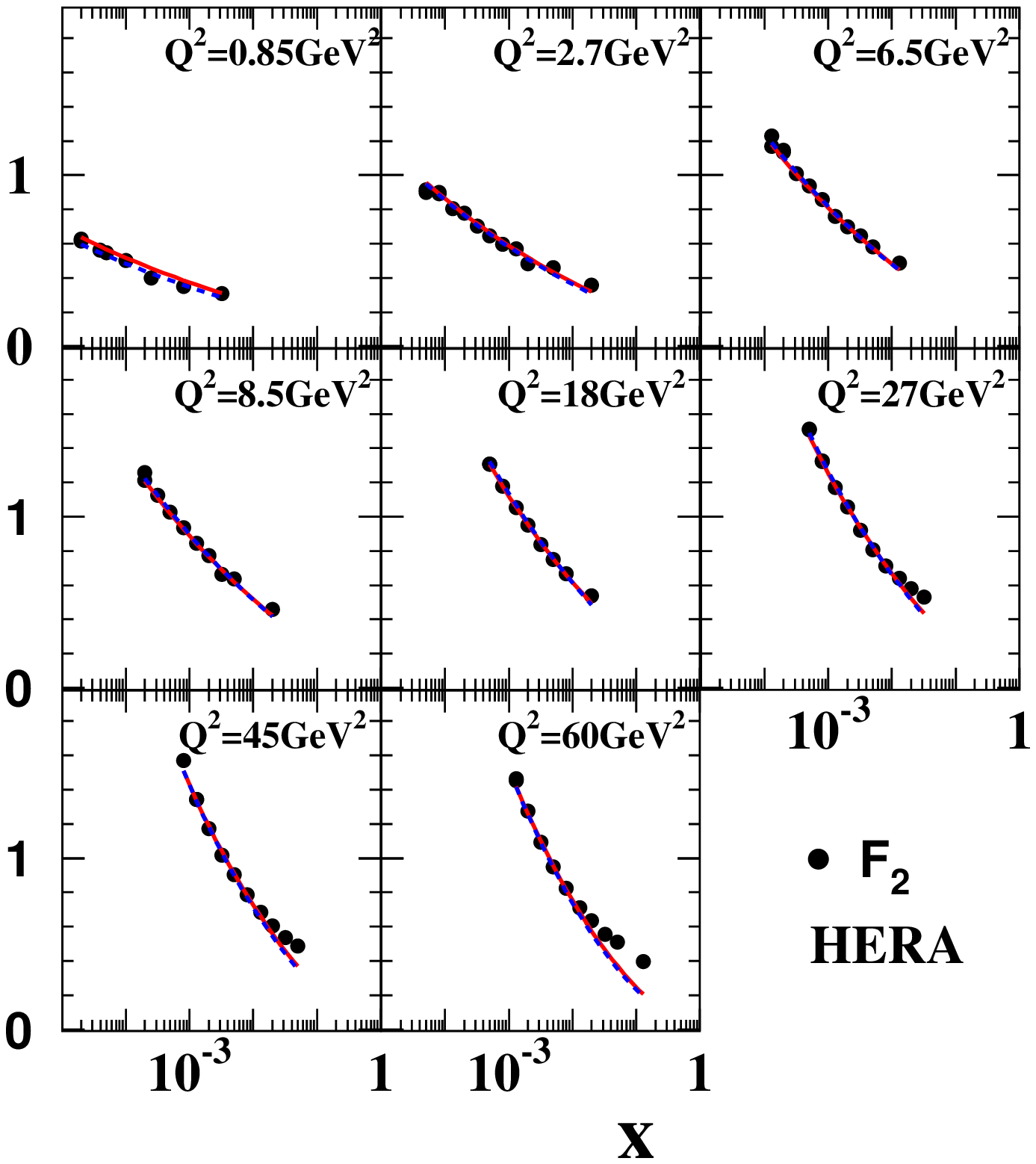}  
      \caption{Our fit of $F_2$ with the values of  parameters given by Table 1. The first set of parameters is shown in solid red curves while the second in blue dotted  lines. The data is taken from Ref.\cite{HERA1}.}
\label{fit}
   \end{figure}

Table 1 gives the values of the fitting parameters and \fig{fit} demonstrates the quality of the fit.

\begin{figure}[h]
    \centering
  \leavevmode
      \includegraphics[width=10cm]{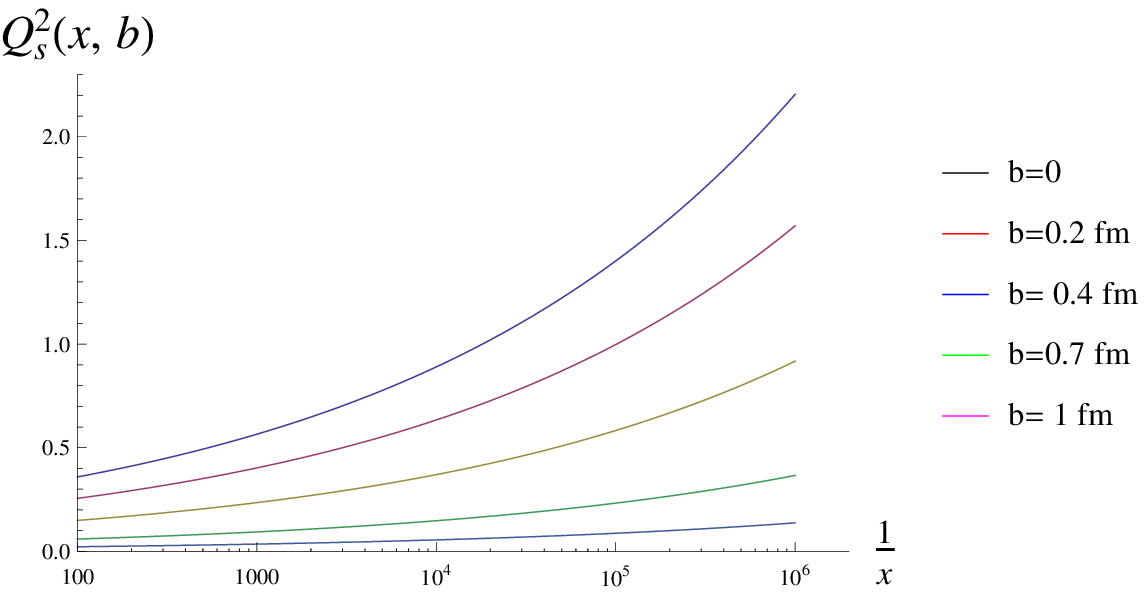}  
      \caption{The value of the saturation momentum $Q^2_s\Lb x, b\Rb$ versus $x$ at fixed $b$ for the  parameters given by Table 1. }
\label{qs}
   \end{figure}

One can see from the Table 1 that both  sets of parameters give good descriptions of the experimental data and the values of the parameters are very close for both fits. The value of $\lambda$ leads to very small  value of the QCD coupling $\bas$ in \eq{LAM},  but, as we have discussed, the estimates in the NLO of perturbative QCD lead the value of $\lambda$\cite{MUT,KMRS} which is  much smaller than in \eq{LAM} and which is close to the value that we obtain from the fit. The value of $N_0$ turns out to be small enough,  to use our matching procedure.  $Q^2_0 = m^2 \,x^\lambda_0$ looks reasonable  and generates the values of the saturation momentum shown in \fig{qs}. 

\begin{figure}[h]
    \centering
  \leavevmode
      \includegraphics[width=10cm]{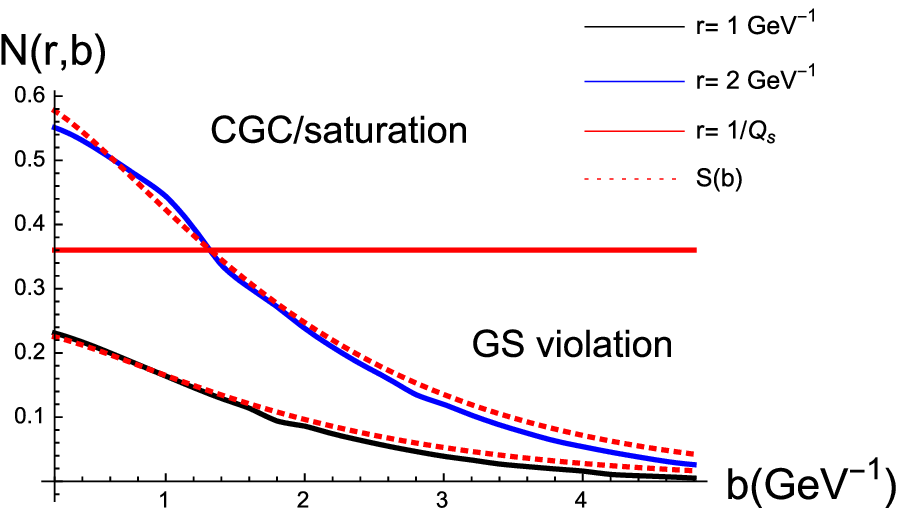}  
      \caption{The  $b$-dependence of the scattering amplitude for the  parameters given by Table 1. $S\Lb b\Rb$ is given by \protect\eq{QSB}.}
\label{nb}
   \end{figure}

The value of $m$ is smaller than the typical mass in the electro-magnetic form factor of the proton, but we do not expect that it will be the same. The mass that we extract from our fit is close to the mass of $\rho$-meson. However,  it should be noted that the decrease of $Q^2_s$ at large $b$ is proportional to $\exp\Lb - \frac{m}{1 - \gamma_{cr}}\,  b\Rb = \exp\Lb - 1.2 \Lb GeV^{-1}\Rb\,b\Rb$. On the other hand the  behavior of amplitude  with $b$ differs from 
 the saturation scale. In \fig{nb} one can see that both the saturation,  and the violation of the geometric scaling behavior influence the resulting b-dependence of scattering amplitude. The saturation flattens the $ b$-dependence at small values of $b$, while the large $b$ behaviour shows a more rapid decrease than the $b$-dependence of the saturation scale  (see \fig{nb}).

It should be stressed that in the framework of our parametrization of the $b$-dependence of the saturation momentum, the scattering amplitude decreases as $\exp\Lb - m b\Rb$ while in all other models on the market it has a Gaussian behavior: $\exp\Lb - m^2\,b^2\Rb$.
     \begin{figure}[h]
    \centering
  \leavevmode
      \includegraphics[width=14cm]{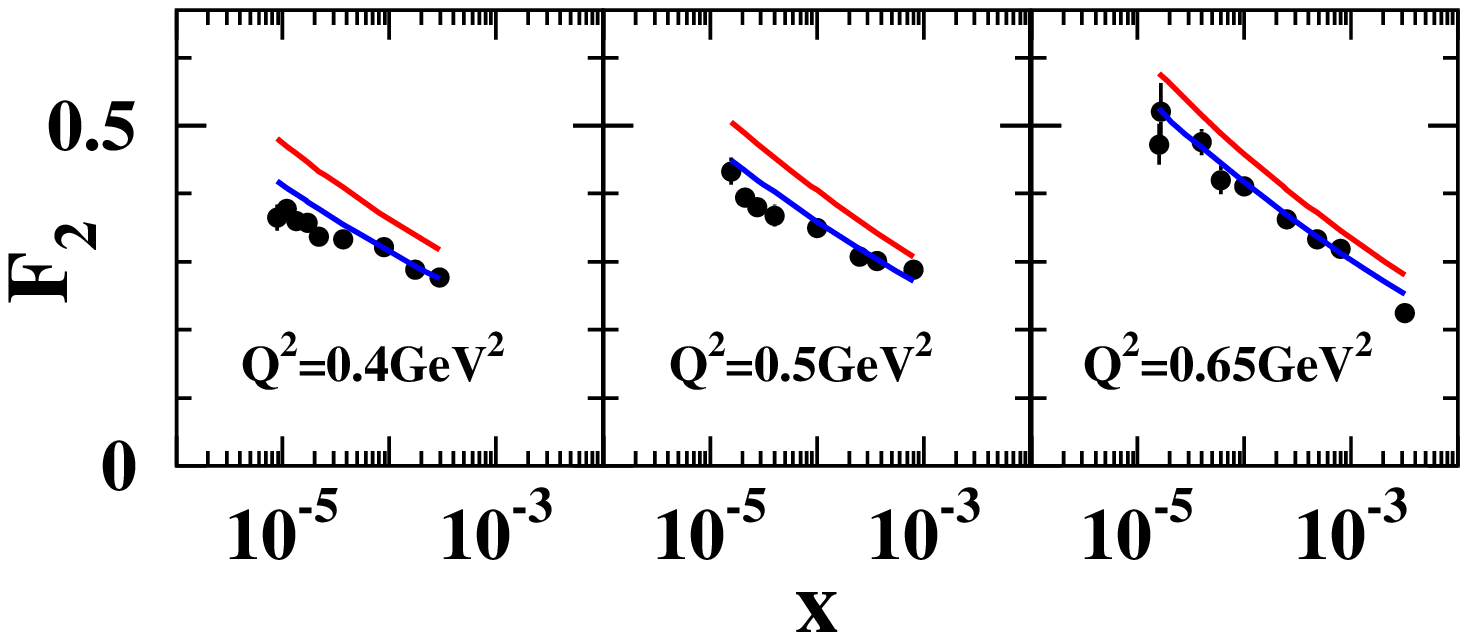}  
      \caption{The  $x$-dependence of $F^{c\bar{c}}_2$ at small  values of $Q^2 \,<\,0.85\,GeV^2$ for the  parameters given by Table 1. The red (upper)  line corresponds to set 1(upper row  of  Table 1)  while  the blue one (low)  is the description with set 2. The data are taken from Ref.\protect\cite{HERAFL1,HERAFL2}.}
\label{f2sq}
   \end{figure}

 
   \fig{f2sq} we present the comparison between our fit of $F_2$ with two sets of parameters at low values of $Q$.
The set with large masses of quarks leads to a much better description illustrating the the non-perturbative corrections to the wave function of the virtual photon are essential at $Q^2 \,<\,0.85 \,GeV^2$.

  \section{Comparison with the experimental data}
  In this section we compare the predictions of our approach using the set of parameters  shown in Table 1, with the experimental data.
  
\begin{figure}[h]
    \centering
  \leavevmode
      \includegraphics[width=14cm]{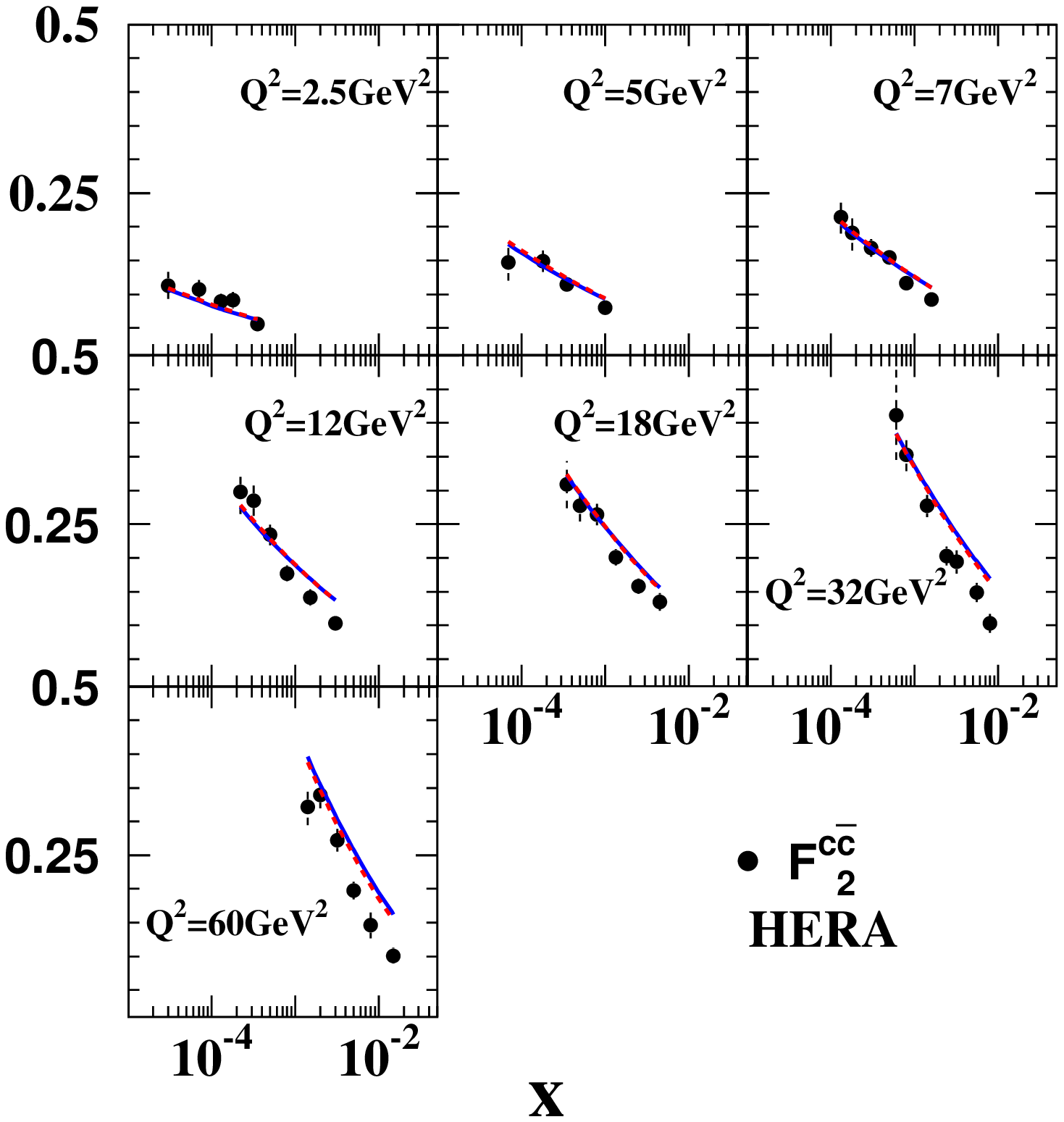}  
      \caption{The  $x$-dependence of $F^{c\bar{c}}_2$ at fixed values of $Q^2$: $ 0.85\,\leq\, Q^2\,\leq\,60\, GeV^2$ for the  parameters given in Table 1. The data are taken from Ref. \protect\cite{HERA2}.}
\label{cc}
   \end{figure}

 $ \mathbf{F_2^{c\bar{c}} } $ : The contribution of the $ c \bar{c}$ pair to the deep inelastic structure function can be calculated with the same theoretical accuracy as the inclusive $F_2$. In \fig{cc} we compare the HERA data on $F^{cc}_2$  \cite{HERA2} with the theoretical predictions.
 One can see that the agreement is reasonable.
   
 ~

\begin{figure}[h]
    \centering
  \leavevmode
      \includegraphics[width=14cm]{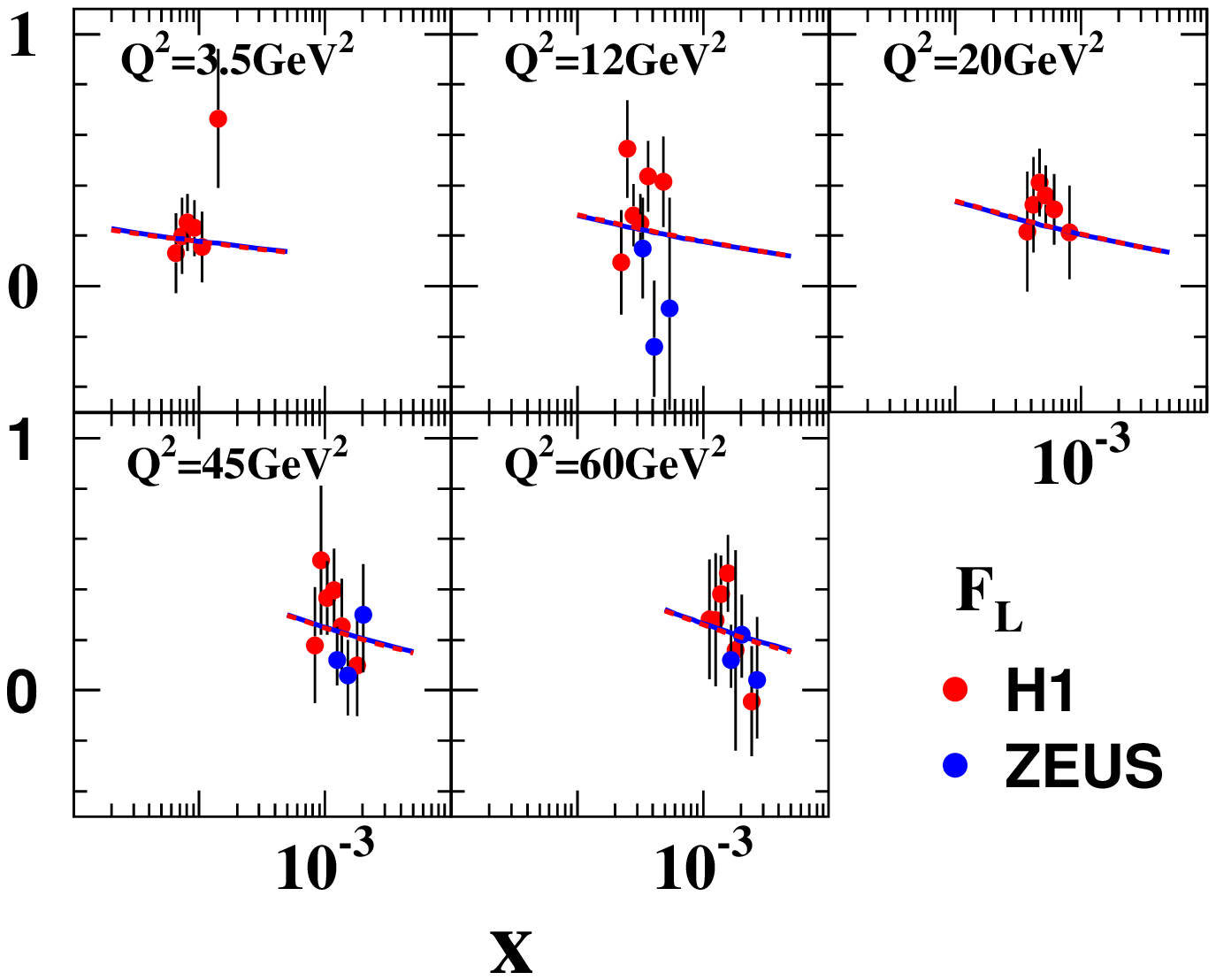}  
      \caption{The  $x$-dependence of $F_L$ at fixed values of $Q^2$: $ 0.85\,\leq\, Q^2\,\leq\,60\, GeV^2$ for the  parameters given in Table 1. The red (blue) lines correspond to set 1 and set 2 fits. The data are taken from Ref. \protect\cite{HERAFL2}.}
\label{f2l}
   \end{figure}

   $ \mathbf{F_L } $ :  $F_L$ can be calculated within the same accuracy as  $F^{c\bar{c}}_2$,  and the comparison with the scant data  available \cite{HERAFL1,HERAFL2} is plotted in \fig{f2l}. Two sets produce  the same quality of the descriptions since the values of $Q$ are rather large.

  {\bf Main formulae for cross sections of  exclusive diffractive production:}
   
   First, we need  first to calculate the scattering amplitude at fixed $t$ which takes the following form \cite{SATMOD6}:
\begin{eqnarray}
  \mathcal{A}^{\gamma^* p\rightarrow V p}_{T,L}(x,Q,\Delta) = \mathrm{i}\,\int\!\dif^2\vec{r}\int_0^1\!\frac{\dif{z}}{4\pi}\int\!\dif^2\vec{b}\;(\Psi_{V}^{*}\Psi)_{T,L}\; 
  \mathrm{e}^{-\mathrm{i}[\vec{b}-(1-z)\vec{r}]\cdot\vec{\Delta}}
  \; N\Lb r, Y; b\Rb
  \label{SA}
\end{eqnarray}
   where $|\vec{\Delta}  |^2 = - t$. %
   
The elastic diffractive cross section is then given by
\begin{eqnarray}
  \frac{\dif\sigma^{\gamma^* p\rightarrow V p}_{T,L}}{\dif t}
  =\frac{1}{16\pi}\left\lvert\mathcal{A}^{\gamma^* p\rightarrow V p}_{T,L}\right\rvert^2
  =\frac{1}{16\pi}
  \left\lvert
  \int\!\dif^2\vec{r}\int_0^1\!\frac{\dif{z}}{4\pi}\int\!\dif^2\vec{b}\;(\Psi_{V}^{*}\Psi)_{T,L}\; 
  \mathrm{e}^{-\mathrm{i}[\vec{b}-(1-z)\vec{r}]\cdot\vec{\Delta}}
  \; N\Lb r, Y; b\Rb
  \right\rvert^2.
  \label{DXS}
\end{eqnarray}
If  we restrict ourselves to measuring two integrated observables:  $\sigma^{\gamma^* p\rightarrow V p}_{T,L}  = \int d t   \frac{\dif\sigma^{\gamma^* p\rightarrow V p}_{T,L}}{\dif t}$ and $B_D = \ln\Lb \frac{\dif\sigma^{\gamma^* p\rightarrow V p}_{T,L}}{\dif t}\Big{/} \frac{\dif\sigma^{\gamma^* p\rightarrow V p}_{T,L}}{\dif t}|_{t = 0}\Rb$,
\eq{SA} and \eq{DXS} can be simplified.

 The formulae for them take the following form
\bea
&&\sigma^{\gamma^* p \to V p}\,=\, \frac{1}{16\pi}
\Bigg(\int \!\dif^2\vec{b} \label{TXS}\\
&&
\Big\{ \int\!\dif^2\vec{r}\int_0^1\!\frac{\dif{z}}{4\pi}(\Psi_{V}^{*}\Psi)_{T,L}\; 
  \; N\Lb r, Y; b\Rb\Big\} \,\Big\{ \int\!\dif^2\vec{r'}\int_0^1\!\frac{\dif{z'}}{4\pi}(\Psi_{V}^{*}\Psi)_{T,L}\; 
  \; N\Lb r', Y; \vec{b} - (1-z')\vec{r'}\Rb\Big\}^* \Bigg);\nn\\
  &&\sigma^{\gamma^* p \to V p}\,B_D\,=\, \frac{1}{16\pi}
\Bigg(\int \!\dif^2\vec{b}\,\dif^2\vec{r}\,\dif^2\vec{r'} \int_0^1\!\frac{\dif{z}}{4\pi}\int_0^1\!\frac{\dif{z'}}{4\pi} \label{BD}\\
&& \Lb 4 b^2 + (1 - z)^2 r^2 + (1 - z')^2\,r'^2\Rb\Big\{(\Psi_{V}^{*}\Psi)_{T,L}\; 
  \; N\Lb r, Y; b\Rb\Big\} \,\Big\{ (\Psi_{V}^{*}\Psi)_{T,L}\; 
  \; N\Lb r', Y; \vec{b} - (1-z)\vec{r'}\Rb\Big\}^* \Bigg).\nn  \eea

   {\bf Deeply virtual compton scattering(DVCS) }  
   
\begin{figure}[h]
    \centering
  \leavevmode
  \begin{tabular}{ccc}
      \includegraphics[width=6cm]{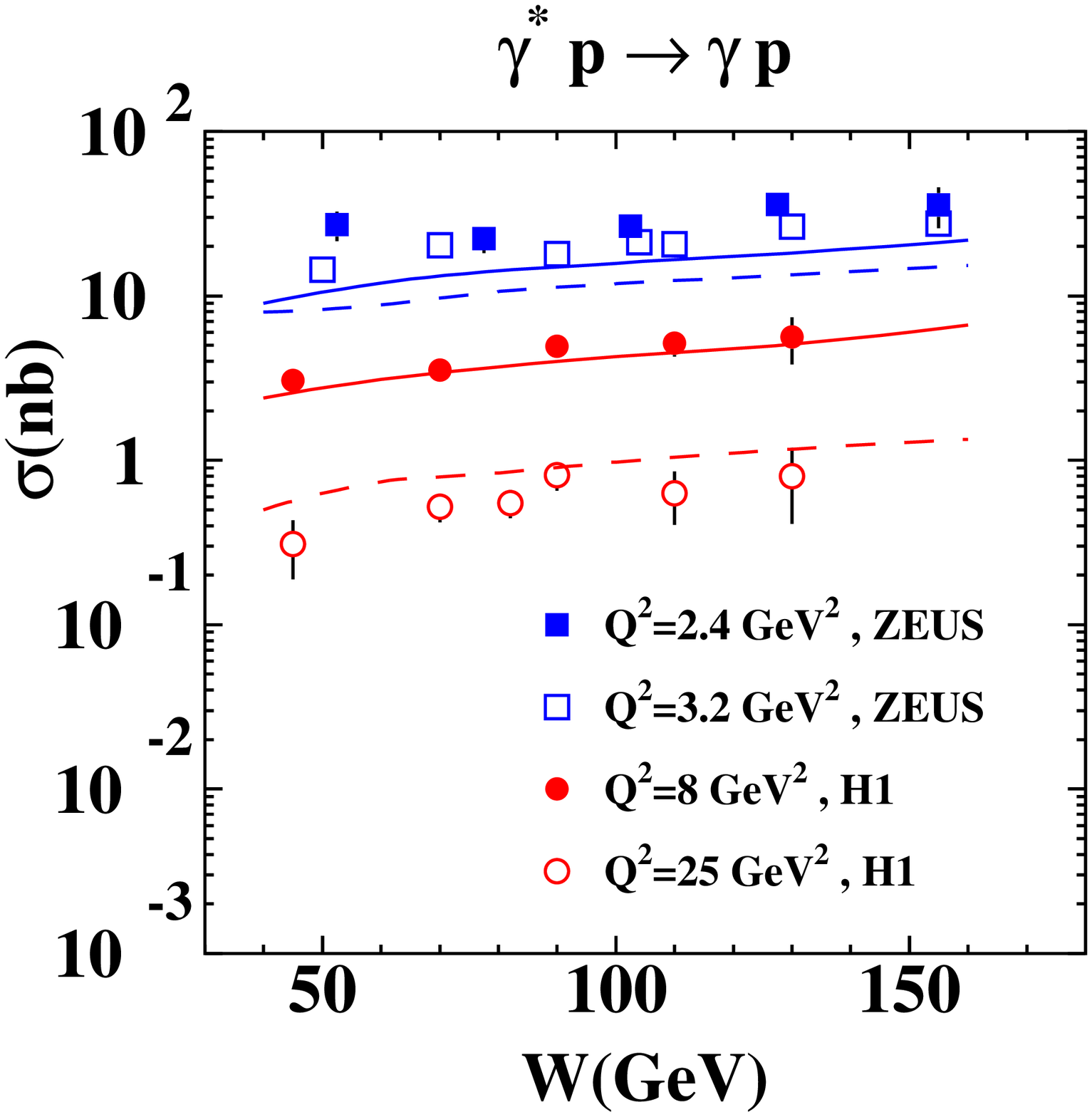}  &    \includegraphics[width=6cm]{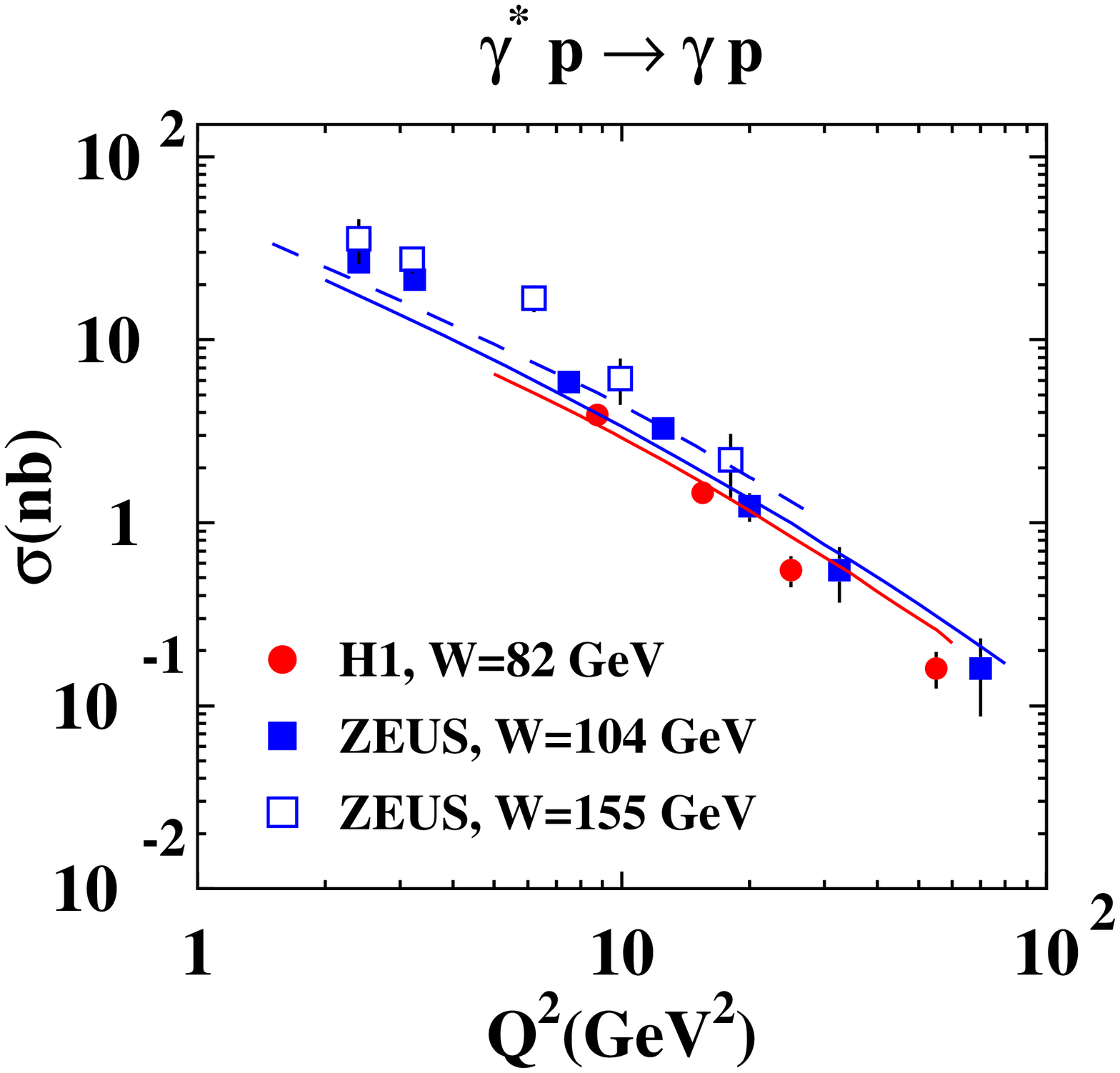}   & \includegraphics[width=6cm]{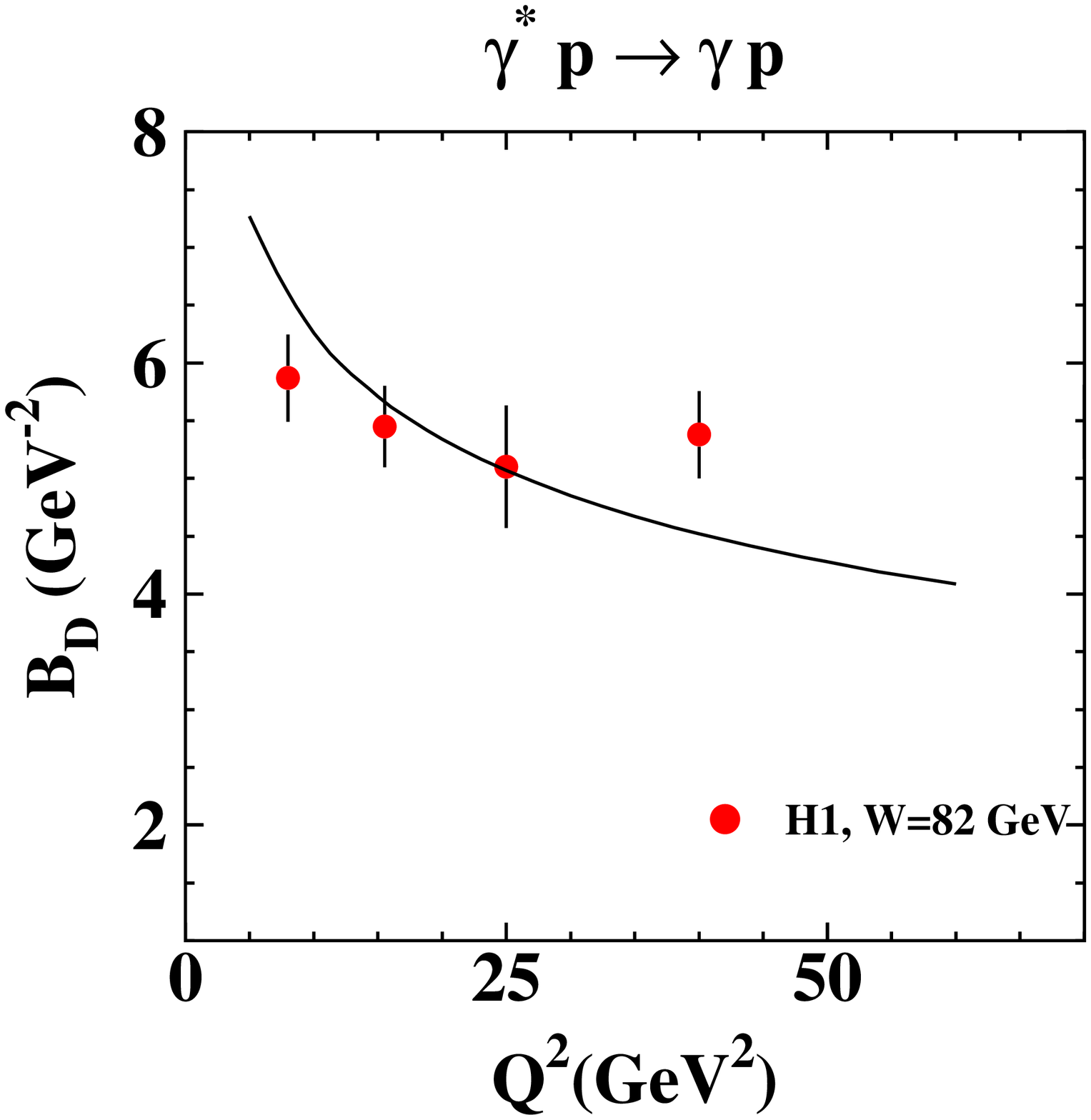}\\
      \fig{dvcs}-a&\fig{dvcs}-b&\fig{dvcs}-c\\
      \end{tabular}
       \caption{ The energy (\fig{dvcs}-a) and $Q^2$   (\fig{dvcs}-b)  dependence  of the cross section for deep inelastic compton scattering. \fig{dvcs}-c shows the energy dependence of the slope. The data are taken from \protect\cite{HERAVIRTCOM}. The lines correspond to the set 1 of the parameters.  The red color describe the H1 data while the blue one stand for ZEUS data.}
\label{dvcs}
   \end{figure}

   Deeply inelastic Compton scattering: $\gamma^* + p \to \gamma + p$, can be calculated to  the same accuracy as all
   reactions that has been discussed,  but it suffers from the errors due to the procedure for calculating both the real part of the amplitude and its skewness.   \fig{dvcs} shows the comparison with the data, taken from Refs.\cite{HERAVIRTCOM}.
One  can see that the description of the cross sections are satisfactory, while the energy dependence of the slope turns out to be more complicated  than  in the experimental data.
   
   ~

   {\bf Total cross sections for diffractive production of vector mesons $\mathbf{\sigma^{\gamma^* p \to V p}}$ (see   \eq{TXS})}
   
\begin{figure}[h]
    \centering
  \leavevmode
  \begin{tabular}{ccc}
      \includegraphics[width=6cm]{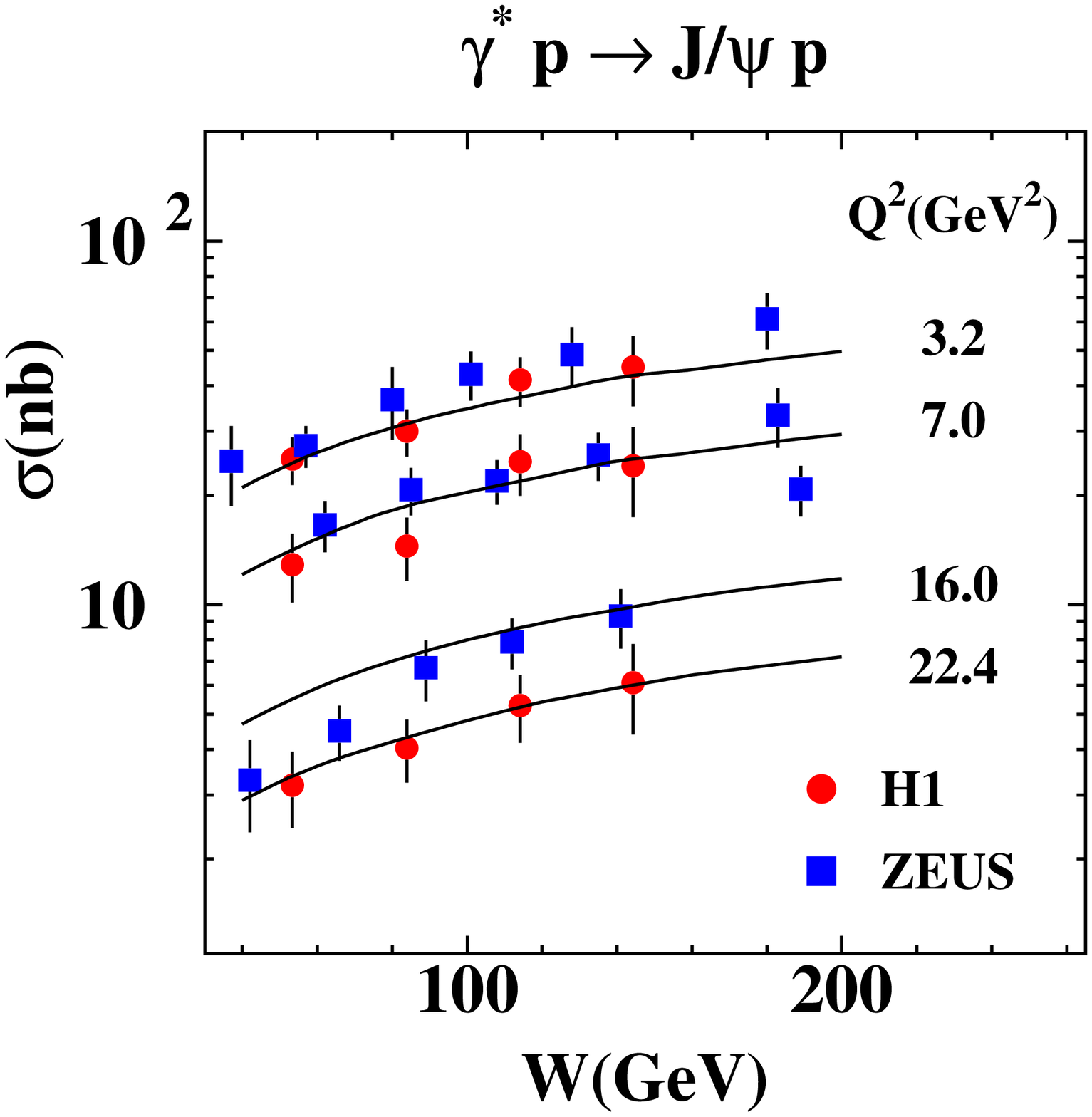}  &    \includegraphics[width=6cm]{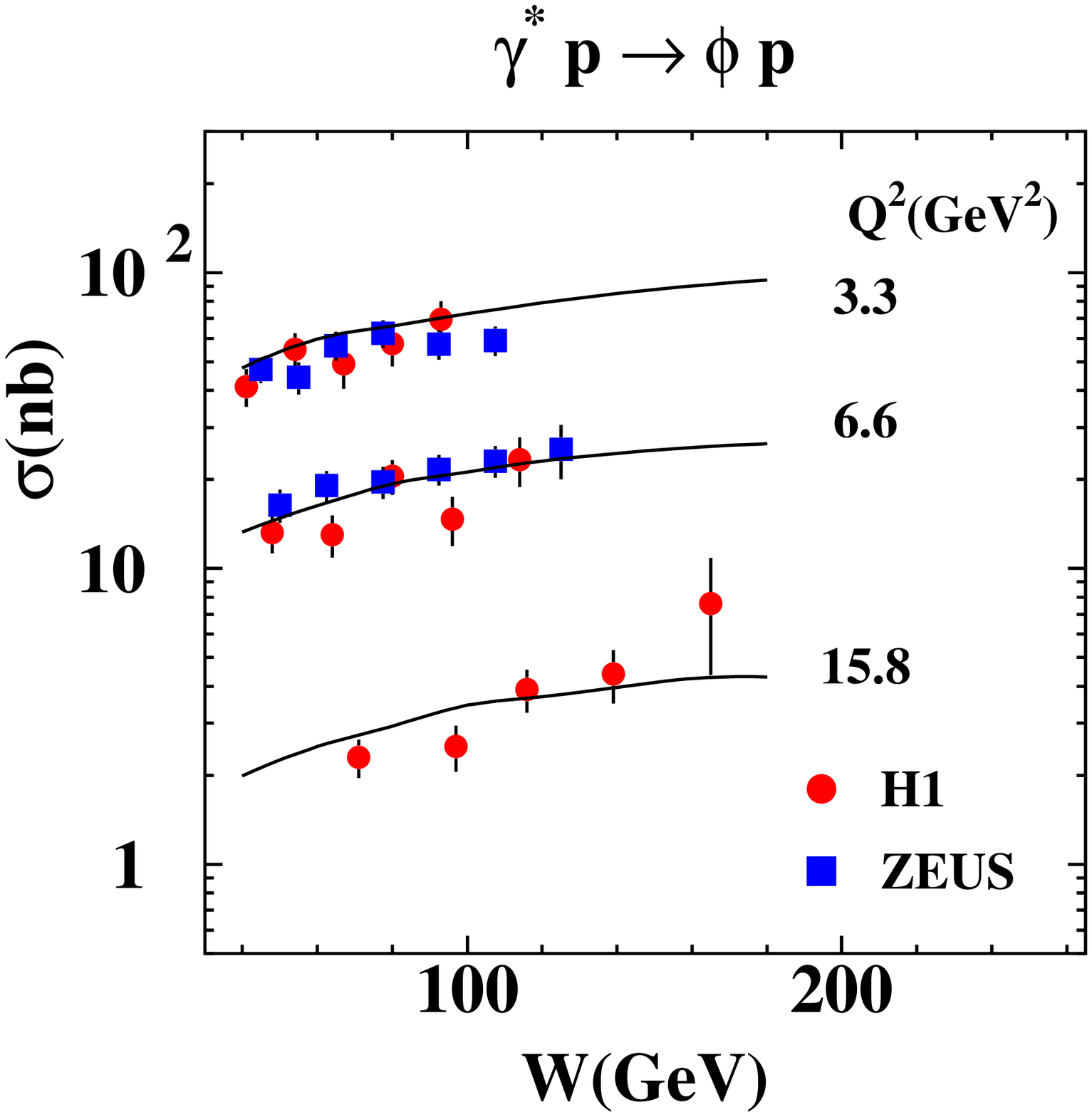}   & \includegraphics[width=6cm]{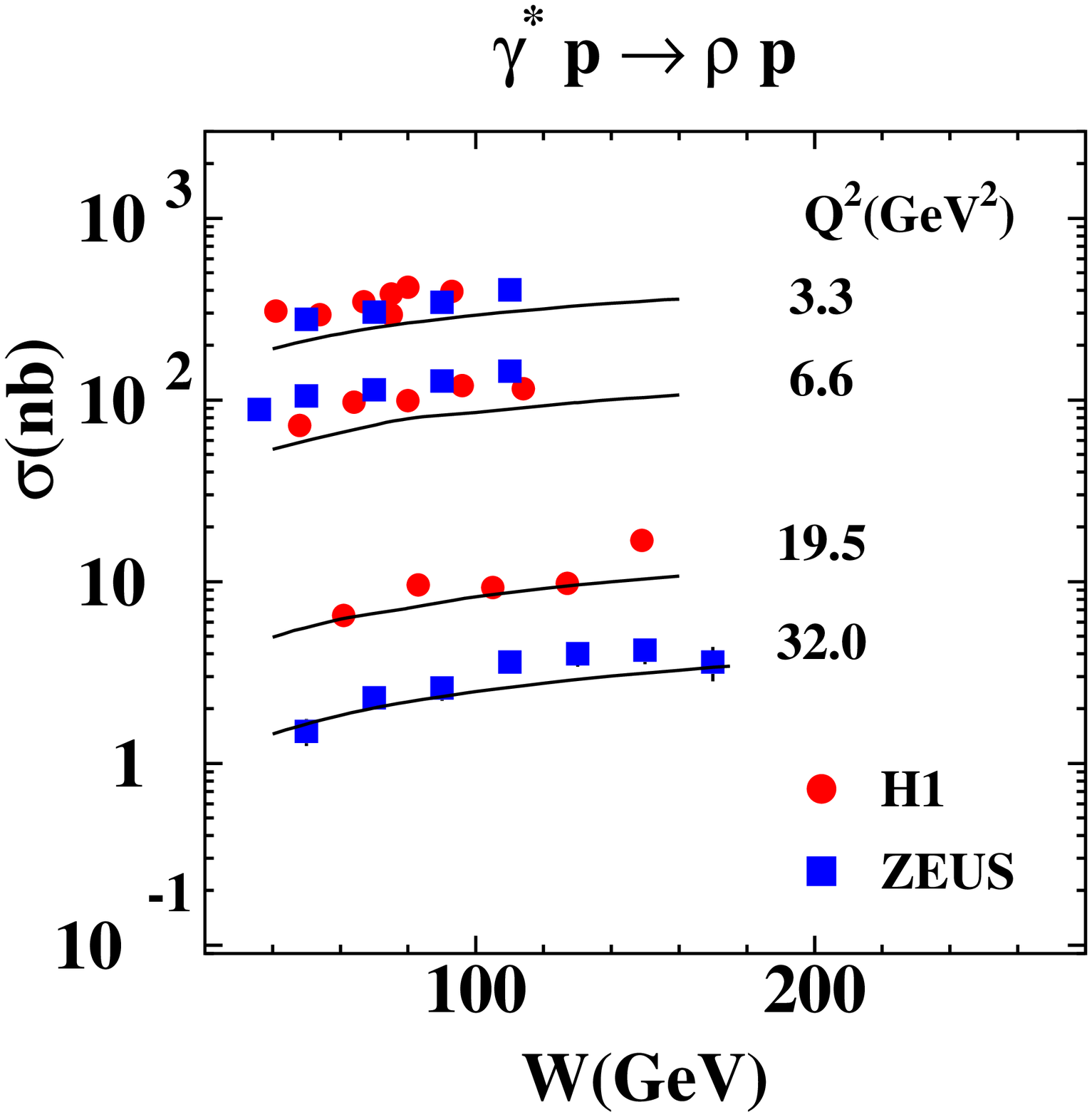}\\
      \fig{txsw}-a&\fig{txsw}-b&\fig{txsw}-c\\
      \end{tabular}
       \caption{The  $W$-dependence of $\sigma^{\gamma^* p \to V p}$   for vector meson production at fixed values of $Q^2$. The data are taken from \protect\cite{HERASLOPE,HERAPSI,HERAH1PSISL,HERAZPHISL}.}
\label{txsw}
   \end{figure}

\begin{figure}[h]
    \centering
  \leavevmode
  \begin{tabular}{ccc}
      \includegraphics[width=6cm]{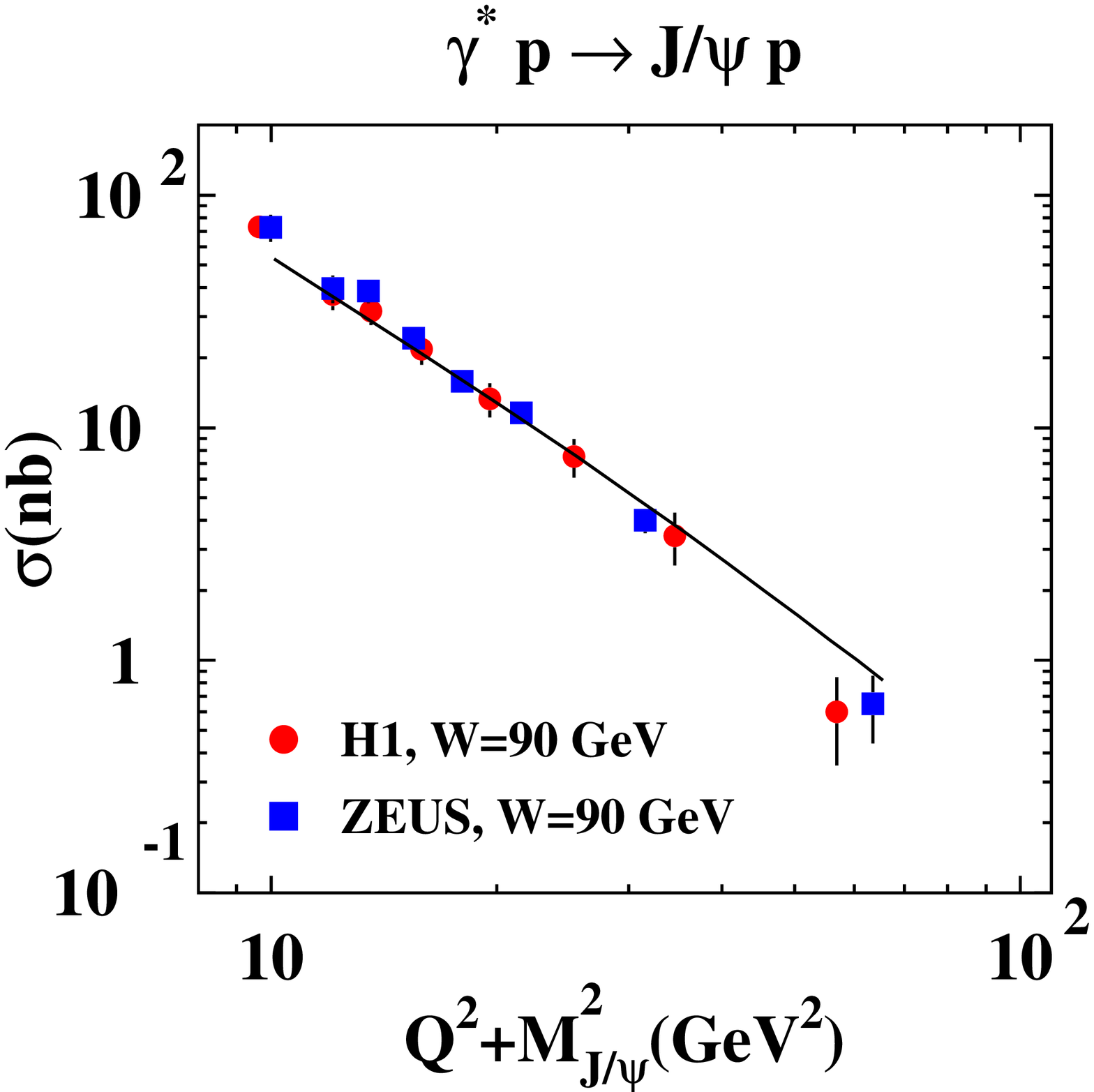}  &    \includegraphics[width=6cm]{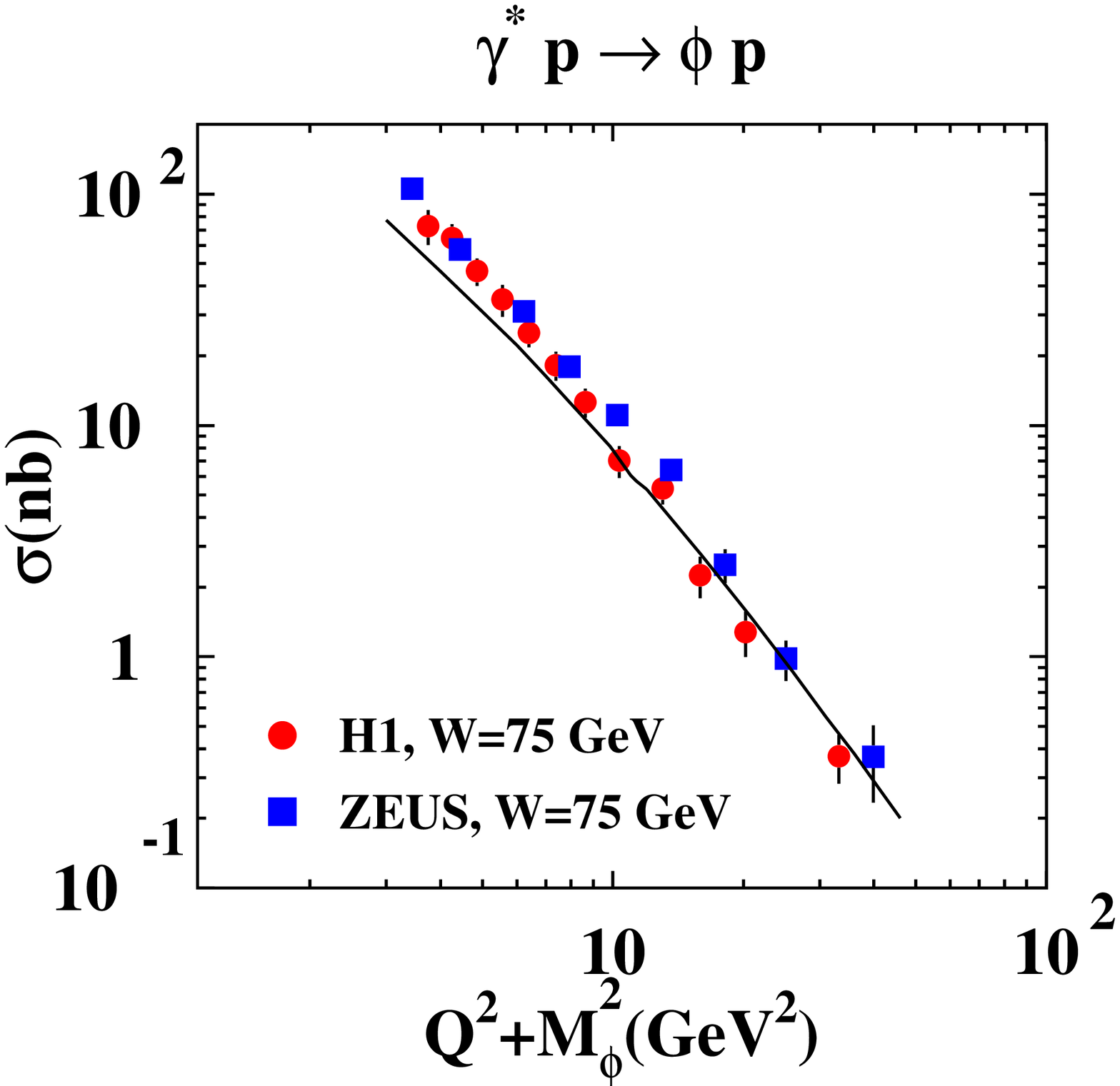}   & \includegraphics[width=6cm]{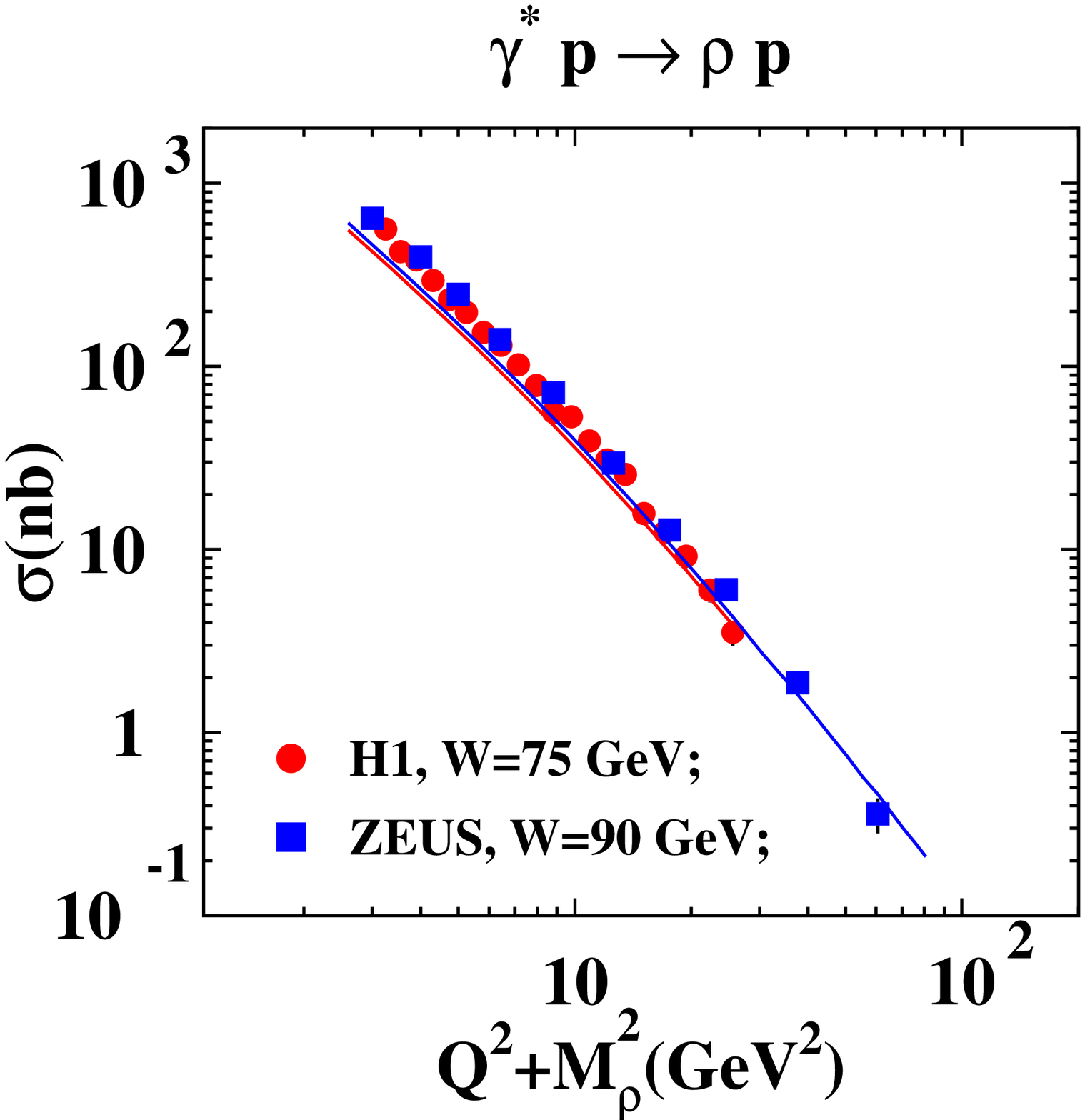}\\
      \fig{txsq}-a&\fig{txsq}-b&\fig{txsq}-c\\
      \end{tabular}
       \caption{The  $Q^2 + M^2_V$-dependence of $\sigma^{\gamma^* p \to V p}$   for vector meson production at fixed values of W. The data are taken from Refs.\protect\cite{HERASLOPE,HERAPSI,HERAH1PSISL,HERAZPHISL}.}
\label{txsq}
   \end{figure}

   In \fig{txsw} and \fig{txsq} we  compare our prediction  with the experimental data \cite{HERASLOPE,HERAPSI,HERAH1PSISL,HERAZPHISL}
   for productions of  J/$\Psi$, $\phi$ and $\rho$-mesons.

      We introduce the factor $(1 + \rho^2) R^2_G$ to account for the real part of the amplitude and for the skewness as has been discussed in section 2.4. One can see that agreement is good,  and the heavier the produced meson, the better the agreement, as  expected.
      
\begin{figure}[h]
    \centering
  \leavevmode
  \begin{tabular}{ccc}
      \includegraphics[width=6cm]{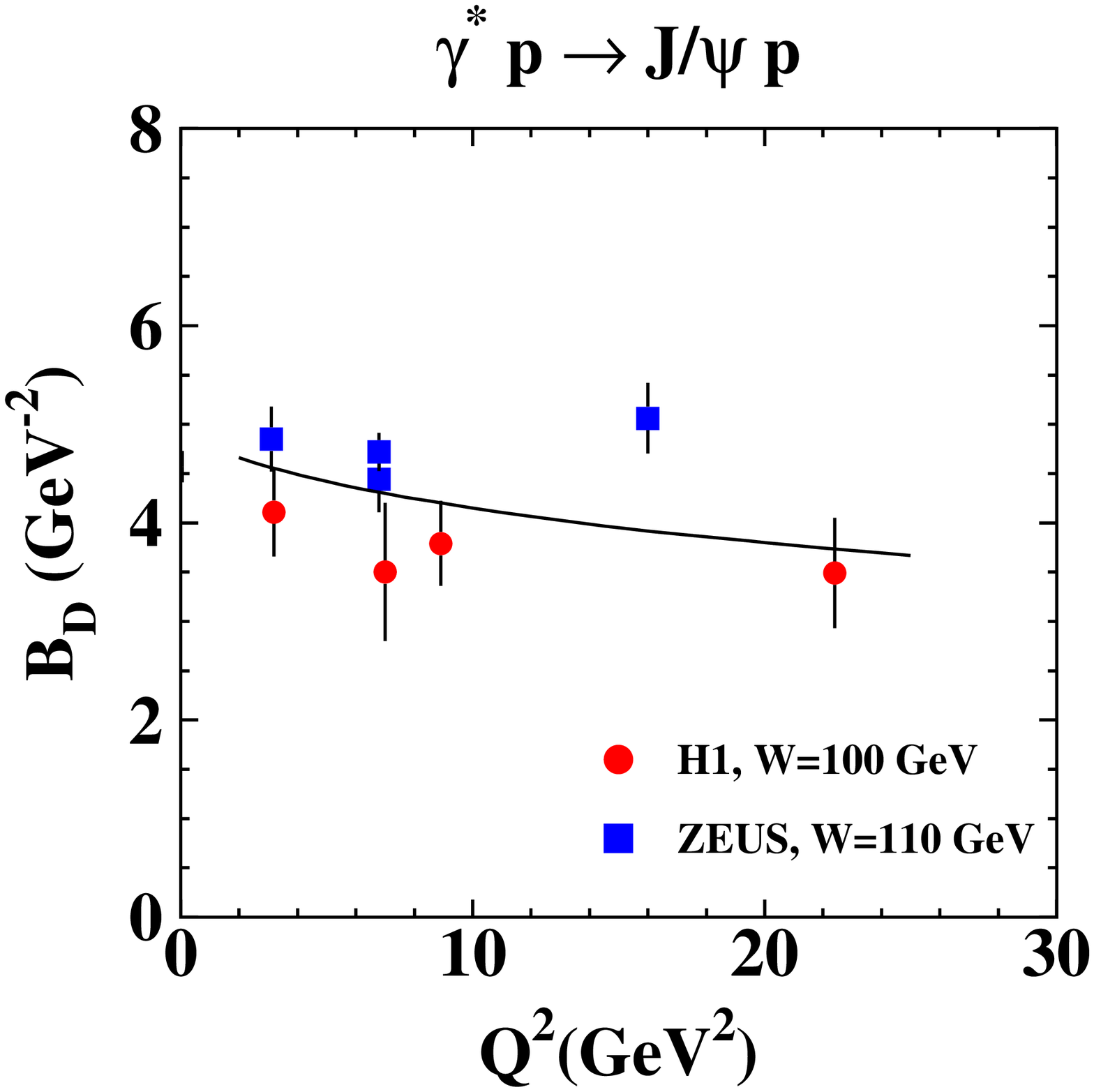}  &    \includegraphics[width=6cm]{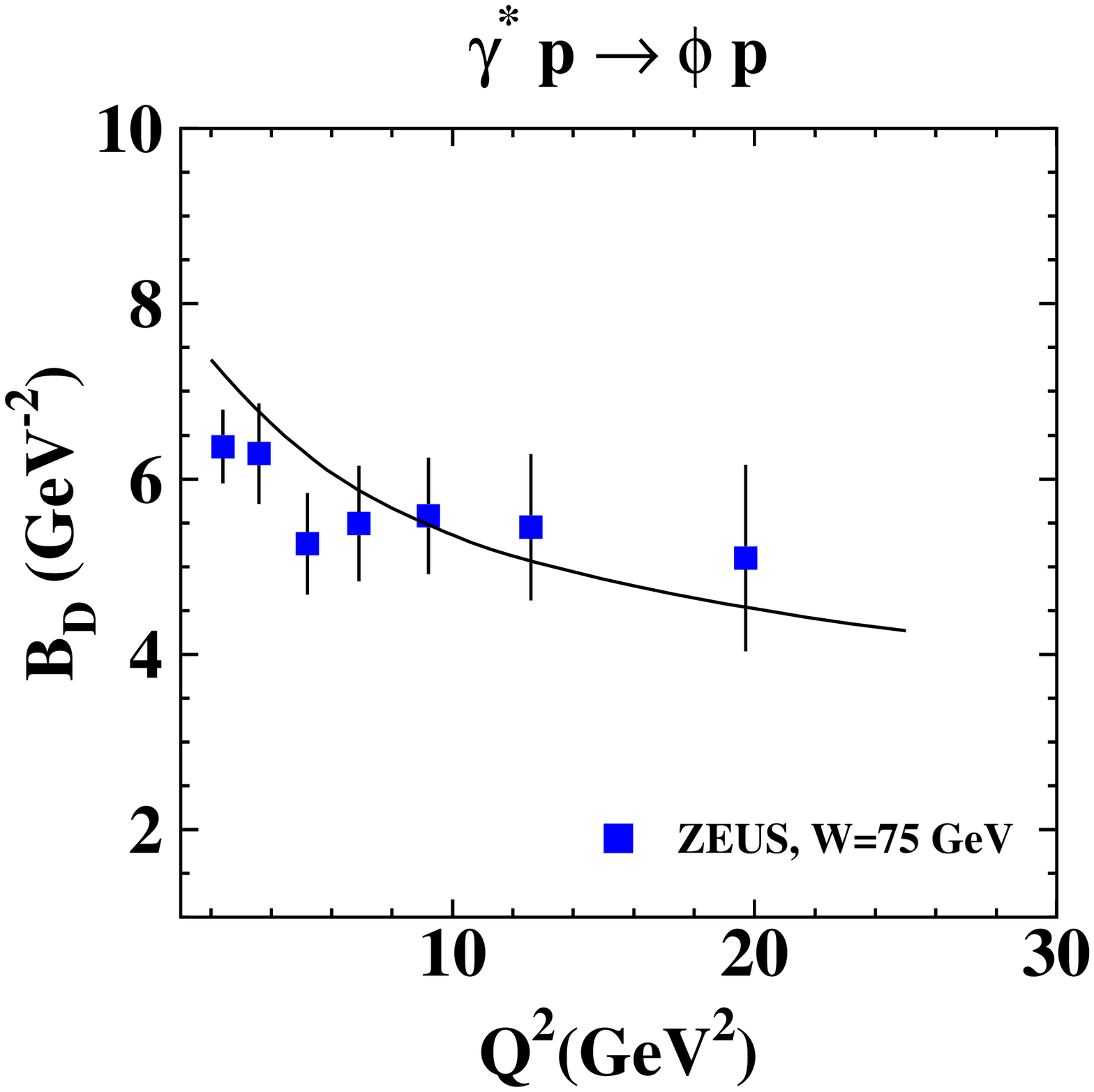}   & \includegraphics[width=6cm]{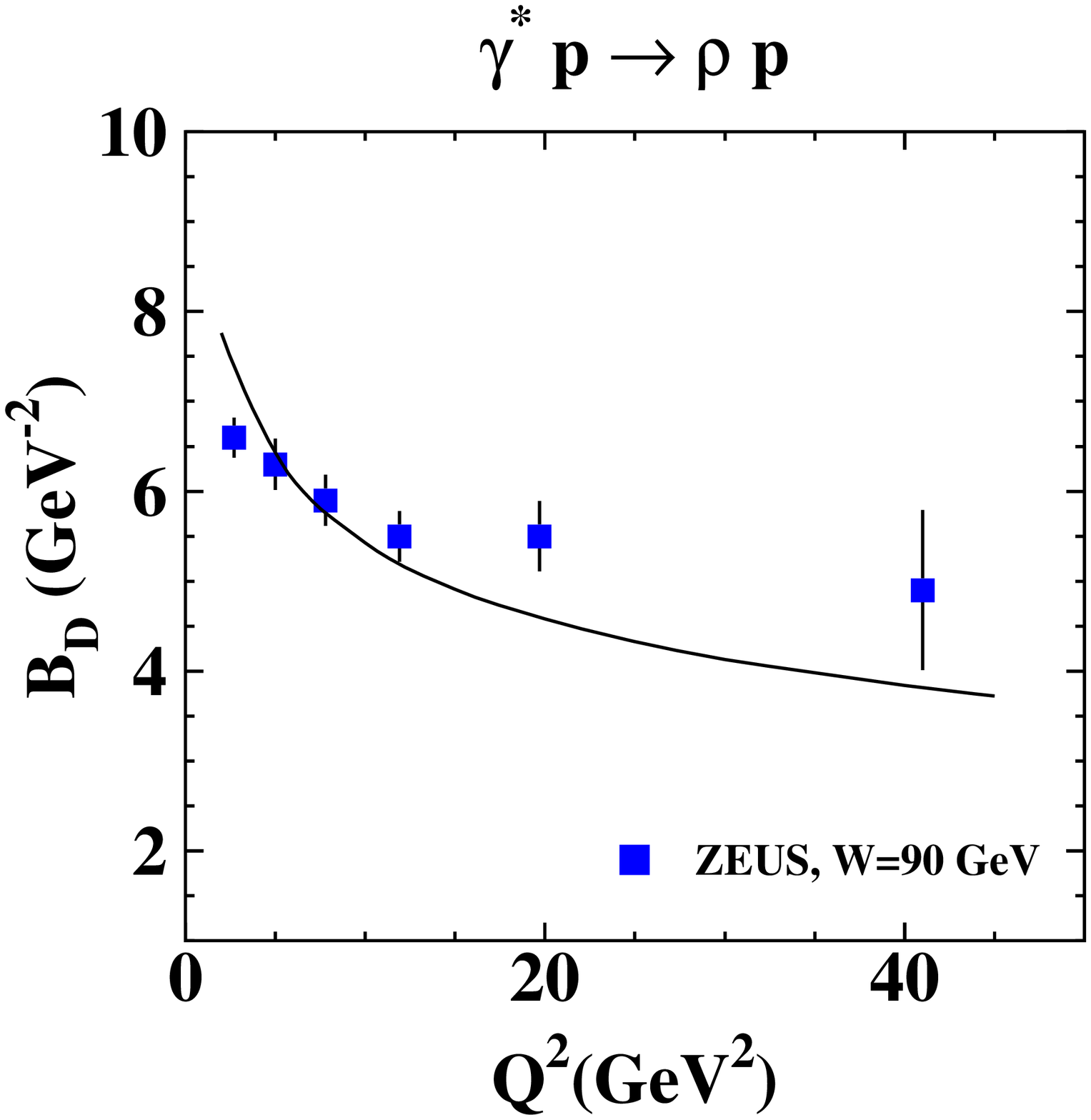}\\
      \fig{bd}-a&\fig{bd}-b&\fig{bd}-c\\
      \end{tabular}
       \caption{The  $Q^2 $-dependence of $B_D$   for vector meson production at fixed values of W. The data are taken from Refs.\protect\cite{HERASLOPE,HERAPSI,HERAH1PSISL,HERAZPHISL}.}
\label{bd}
   \end{figure}

\begin{figure}[h]
    \centering
  \leavevmode
  \begin{center}
      \includegraphics[width=9cm]{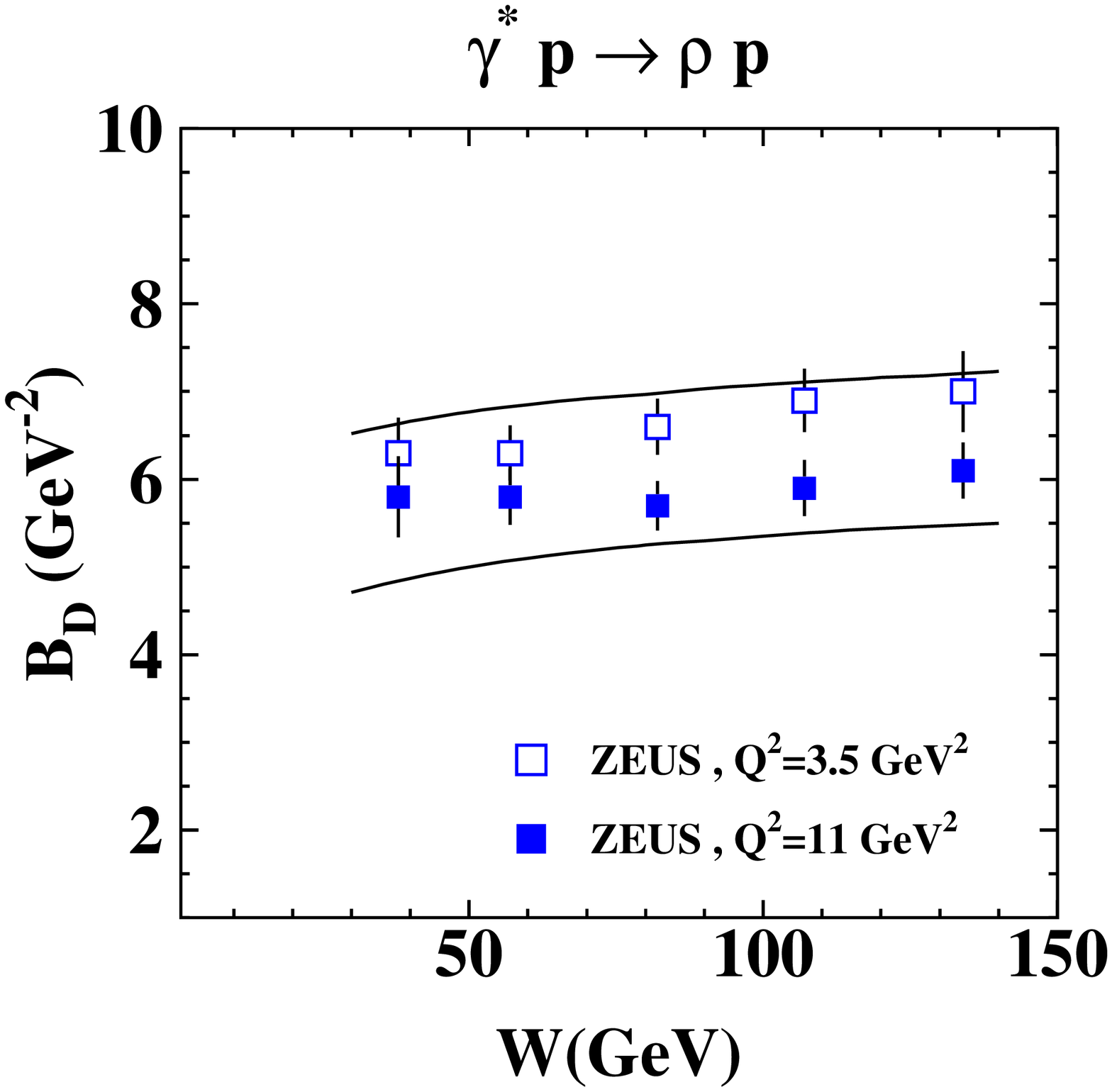}  
      \end{center}
             \caption{The  $W$-dependence of $B_D$   for $\rho$ production at fixed values of $Q^2$. The data are taken from Refs.\protect\cite{HERASLOPE,HERAPSI,HERAH1PSISL,HERAZPHISL}.}
\label{bdw}
   \end{figure}

      {\bf The slopes $\mathbf{B_D}$:} $B_D$ are calculated using \eq{BD} and \fig{bd} shows the comparison with the experiment data taken from Refs. \cite{HERASLOPE,HERAPSI,HERAH1PSISL,HERAZPHISL}. In \fig{bdw} we plot the dependence of the slope on the energy W. We  are able to describe  the values and main regularities of  the slope's behavior on $Q^2$ and $W$. \fig{bdw} shows the the shrinkage of the diffraction peak is very mild.

      ~
      
      ~

      ~

      ~
      
       
       \section{Conclusions}
       
 In this paper we add to the  many CGC/saturation models          \cite{SATMOD0,SATMOD1,BKL,SATMOD2,IIM,SATMOD3,SATMOD4,SATMOD5,SATMOD6,SATMOD7,SATMOD8,SATMOD9,SATMOD10, SATMOD11,SATMOD12,SATMOD13,SATMOD14,SATMOD15,SATMOD16,SATMOD17}    two crucial ingredients:  the correct solution to the non-linear (BK) equation \cite{BK} in the saturation region, and   impact parameter distribution that leads to
 exponential decrease of the saturation momentum at large impact parameters  and to  power-like decrease at large transferred momentum that follows from perturbative QCD. We obtain the  solution to the BK equation based on the ideas proposed in Ref.\cite{IIM}: to match the analytical solution for the scattering amplitude deep inside of the saturation domain \cite{LETU}   with the solution in the vicinity of the saturation momentum \cite{IIML,MUT}. The novel features of our approach   is that we use the exact form of the solution to the BK equation (see section 2.2 and Refs.\cite{LETU,CLM}) but not the form of   $r^2Q^2_s$ dependence as have bee assumed in all previous attempts.
We introduce  the large impact parameter dependence  through the $b$-dependence of the saturation momentum. The  difference between our and the other attempts  consists in the assumption that $Q^2_s \propto \,\exp\Lb - m b\Rb$ at large $b$,  while in all other models the Gaussian behavior at large $b$\, $ Q^2_s \propto \,\exp\Lb - m ^2\,b^2\Rb$ is assumed. In the framework of our model, the exponential $b$ dependence of $Q^2_s$ translates in the exponential decrease of the scattering amplitude at large $b$,  satisfying the Froissart theorem  \cite{FROI}.
       
       Using four fitting parameters we obtain   good overall agreement with the experimental data.   Comparison with the experimental data we found  two regularities. The first one is , the better we know theoretically the wave functions of the produced mesons, the better the agreement with experimental data.         The second is the fact,  that the energy dependence of the saturation momentum,  is much milder that it is predicted by the BK equation in leading log(1/x) approximation of perturbative QCD.  The last observation, we hope, will stimulate the search for the non-linear equation,  in the next-to-leading order  (NLO) of perturbative QCD. The first attempts to  estimate the energy behavior of the saturation scale in NLO show that the value of $\lambda $ significantly decreases (see Refs.\cite{MUT,KMRS,KOLE,IMMST}).

       We  believe that the model presented here, will be a useful tool to estimate the CGC/saturation effects in the variety of exclusive reactions.
       
       The model includes everything that we know from  theory about  deep inelastic processes and, as such, can be used for the 
      comparison at higher energies including the LHC energies. From a good description of the experimental data, which is not better than in insufficient models of Refs.\cite{SATMOD0,SATMOD1,BKL,SATMOD2,IIM,SATMOD3,SATMOD4,SATMOD5,SATMOD6,SATMOD7,SATMOD8,SATMOD9,SATMOD10, SATMOD11,SATMOD12,SATMOD13,SATMOD14,SATMOD15,SATMOD16,SATMOD17} we can conclude that the CGC/saturation approach does not contradict  available experimental data. The weak statement is nevertheless  stronger than the one that we can make from the models of Refs.   \cite{SATMOD0,SATMOD1,BKL,SATMOD2,IIM,SATMOD3,SATMOD4,SATMOD5,SATMOD6,SATMOD7,SATMOD8,SATMOD9,SATMOD10, SATMOD11,SATMOD12,SATMOD13,SATMOD14,SATMOD15,SATMOD16,SATMOD17}:  shadowing corrections are, perhaps, needed  to describe the experimental data.  Formulating the result of the paper in one sentence, we firmly believe that only our model is reliable  for  extrapolation to higher energy including the LHC one.
      
 We need to discuss why we use a model while, at first sight, we have a good description of the experimental data at least for  the deep inelastic scattering, based on the leading order(LO) or even beyond of the leading order (NLO) perturbative QCD evolution equations (see Refs.\cite{IMMST,AAMQS,ALB}).
  Unfortunately,     these equations as we have discussed,  cannot provide reliable predictions for physical observables including $F_2$, while the problem of large impact parameter behaviour of the scattering amplitude would not be solved theoretically and the CGC/saturation equations would have to be be modified to include this behaviour.  The solutions of the CGC/saturation equations that have been discussed, are valid only `` under the approximation that the dipole scattering amplitude is independent on the impact parameter" (citation from Ref.\cite{ALB}).
  Such an approximation generates  the dipole scattering amplitude $ N  \propto \delta^{(2)}\Lb Q_T\Rb$  and $F_2 \propto \mbox{Im} N\Lb Q_T=0\Rb\,\to \,\infty$,
   the prediction which cannot be compared with the experimental data. Therefore, we doomed to build models for comparing with the experimental data.  Using the results of these models we can evaluate how  essential is the $b$-dependence for the description of the experimental data, and see that it is essential even in the HERA kinematic region, and will be even more important at higher energy (i.e. at the LHC).
  
  The second question is why we improve the model based on the leading order perturbative QCD while we know that the next-to-leading QCD corrections (NLO) are essential. The answer is: first we need to build a reliable model in the leading order QCD and only after this we can build the model that takes into account the NLO corrections. We believe that the NLO corrections will change all parameters of the model especially the energy behaviour of the saturation scale leading to small values of $\lambda$ as it was demonstrated in Refs.\cite{MUT,KOLE,IMMST}. Since the value of $\lambda$ can be found from the solution of the linear equation we trust these estimates and, as we have discussed, we see in this NLO result the support for the phenomenological value of $\lambda$. However, we believe that we  do not have enough solid theory information to  discuss in the NLO the change in the behaviour of the scattering amplitude inside the saturation region building a model that includes the impact parameter behaviour. The general form of the non-linear equation in the NLO\cite{BAL} shows that we can use
   the approach of Ref.\cite{LETU}  for finding an equation  at  $z \,\gg\,1$ which takes the form
  \beq \label{CON1}
  \frac{\partial N\Lb Y; x,y\Rb}{\partial Y}\,=\,-\,\bas\, \int d^2 z  \Big\{K^{LO}\Lb x,y|z\Rb + \bas\, K^{NLO}\Lb x,y|z\Rb \Big\}  N\Lb Y; x,y\Rb
  \eeq
  $K^{NLO}
  $ turns out to be  such that $ K^{LO}\Lb x,y|z\Rb + \bas\, K^{NLO}\Lb  x,y|z\Rb \,\to \,0$ for $\bas z \sim 1$\cite{KMRS,KOLE,IMMST}. It means that we cannot specify large $z$ asymptotic behaviour using the NLO kernel. We have to re-sum all orders of $(\bas z)^n$  but we do not know a theory proof of the form of the non-linear equation for such a  re-summed kernel. All attempts to write the non-linear evolution equation for the re-summed kernel (e.g. Refs.\cite{KOLE,IMMST}) use the simplified form of the kernel for the BFKL equation taking into account only its leading twist contribution. The correct coefficient from the full BFKL kernel in front of $z^2$ term in the solution at large $z \,\gg\,1$, is much more important for  phenomenology than the NLO corrections.
  Recalling that the running QCD coupling that should be included in such a re-summation, leads to the violation of the geometric scaling behaviour\cite{SALE}, and 
 that the improved double log approximation of Ref.\cite{IMMST} cannot be used in the saturation region\cite{LETU, KMRS,KOLE}, we see that a lot of theory work must be done before we will be able to build a model that will include the impact parameter behaviour and the NLO corrections to the BK equation. We plan building  such model as a natural goal for future work.

  \section{Acknowledgements}
   We thank our colleagues at Tel Aviv university and UTFSM for
 encouraging discussions. Our special thanks go to Asher Gotsman, 
 Alex Kovner and Misha Lublinsky for elucidating discussions on the
 subject of this paper.
   This research was supported by the BSF grant   2012124, by  the
 Fondecyt (Chile) grants 1130549 and  1140842 and by  DGIP/USM grant 11.15.41.

\end{document}